\documentstyle[multicol,eqsecnum,aps,pra,epsf]{revtex}

\begin{document}

\title{The Random Transverse Ising Spin Chain and Random Walks}

\author{Ferenc Igl\'oi}
\address{Research Institute for Solid State Physics, 
H-1525 Budapest, P.O.Box 49, Hungary\\ Institute for Theoretical Physics,
Szeged University, H-6720 Szeged, Hungary\\ }

\author{Heiko Rieger}
\address{HLRZ, Forschungszentrum J\"ulich, 52425 J\"ulich, Germany}

\date{August 27, 1997}

\maketitle

\begin{abstract}
We study the critical and off-critical (Griffiths-McCoy) regions of the
random transverse-field Ising spin chain by analytical and numerical
methods and by phenomenological scaling considerations. Here we extend
previous investigations to surface quantities and to the ferromagnetic
phase. The surface magnetization of the model is shown to be related
to the surviving probability of an adsorbing walk and several critical
exponents are exactly calculated. Analyzing the structure of low energy
excitations we present a phenomenological theory which explains both the
scaling behavior at the critical point and the nature of Griffiths-McCoy
singularities in the off-critical regions. In the numerical part of the
work we used the free-fermion representation of the model and calculated
the critical magnetization profiles, which are found to follow very
accurately the conformal predictions for different boundary conditions.
In the off-critical regions we demonstrated that the Griffiths-McCoy
singularities are characterized by a single, varying exponent, the
value of which is related through duality in the paramagnetic and
ferromagnetic phases.
\end{abstract}



\newcommand{\bc}{\begin{center}}
\newcommand{\ec}{\end{center}}
\newcommand{\be}{\begin{equation}}
\newcommand{\ee}{\end{equation}}
\newcommand{\beqn}{\begin{eqnarray}}
\newcommand{\eeqn}{\end{eqnarray}}

\begin{multicols}{2}
\narrowtext

\section{Introduction}

Magnetic systems with quenched disorder at very low or even vanishing
temperature have attracted a lot of interest recently. In particular
the quantum phase transition occurring in quantum Ising spin glasses
in a transverse field \cite{qsg} and random transverse Ising
ferromagnets \cite{fisher,senthil} turned out to have a number of
surprising features. For instance the presence of quenched disorder
has more pronounced effects on quantum phase transitions than on those
phase transitions, which are driven by thermal fluctuations. For
example in the Griffiths phase, which is at the disordered side of the
critical point, the susceptibility has an essential singularity in
classical systems, whereas in a random quantum system the
corresponding singularity (Griffiths-McCoy singularity
\cite{griffiths,mccoy}) is much stronger, it is in a power-law form.

Many interesting features of random quantum systems can already be
seen in one-dimensional models. After the pioneering work by McCoy and
Wu \cite{mccoywu} and later studies by Shankar and Murphy
\cite{shankar}, Fisher \cite{fisher} has recently performed an
exhaustive study of the critical behavior of the random
transverse-field Ising spin chain. He used a renormalization group (RG)
approach, which he claims becomes exact at the critical point. The
same type of method has later been used for other $1d$ random quantum
problems \cite{hymanyang} and recently some exact results are obtained
through a mapping of the random $XY$-model onto a Dirac equation in
the continuum limit \cite{mckenzie}.

In the present paper we consider the prototype of random quantum systems the
random transverse-field Ising chain defined by the Hamiltonian:
\be
H=-\sum_l J_l \sigma_l^x \sigma_{l+1}^x-\sum_l h_l \sigma_l^z\;.
\label{hamilton}
\ee
Here the $\sigma_l^x$, $\sigma_l^z$ are Pauli matrices at site $l$ and
the $J_l$ exchange couplings and the $h_l$ transverse-fields are
random variables with distributions $\pi(J)$ and $\rho(h)$,
respectively. The Hamiltonian in eq(\ref{hamilton}) is closely related
to the transfer matrix of a classical two-dimensional layered Ising
model, which was first introduced and studied by McCoy and Wu \cite{mccoywu}.

In the following we briefly summarize the existing exact, conjectured
and numerical results on the random transverse-field Ising chain in
eq(\ref{hamilton}). The quantum control-parameter of the model is
given by
\be
\delta={[\ln h]_{av}-[\ln J]_{av} \over \rm{var}[\ln h]+\rm{var}[\ln J]}\;.
\label{delta}
\ee
For $\delta<0$ the system is in the ordered phase with a non-vanishing
average magnetization, whereas the region $\delta>0$ corresponds to
the disordered phase. There is a phase transition in the system at
$\delta=0$ with rather special properties, which differs in several
respects from the usual second-order phase transitions of pure
systems.  One of the most striking phenomena is that some physical
quantities are not self-averaging, which is due to very broad,
logarithmic probability distributions. As a consequence the {\it
typical value} (which is the value on an event with probability one)
and the {\it average value} of such quantities is different. Thus the
critical behavior of the system is primary determined by rare events,
dominating the averaged values of various observables.

The average surface magnetization close to the critical point vanishes as a
power law $m_s \sim \delta^{\beta_s}$, where
\be
\beta_s=1\;,
\label{betas}
\ee
is an exact result by McCoy and Wu \cite{mccoywu}. The average bulk
magnetization is characterized by another exponent $\beta$, which is
conjectured by Fisher in his RG-treatment \cite{fisher}:
\be
\beta=2-\tau\;,
\label{beta}
\ee
where $\tau=(1+\sqrt{5})/2$ is the golden-mean. The average spin-spin
correlation function $G(l)=[\langle\sigma_i^x
\sigma_{i+l}^x\rangle]_{av}$ involves the average correlation length
$\xi$, which diverges at the critical point as $\xi \sim
|\delta|^{-\nu_{\rm av}}$. The RG result by Fisher \cite{fisher} is
\be
\nu_{\rm av}=2\;,
\label{nu}
\ee
On the other hand the typical correlations have a faster decay, since
$\xi_{\rm typ}\sim |\delta| ^{-\nu_{\rm typ}}$ with $\nu_{\rm typ}=1$
\cite{shankar}.

In a quantum system statistics and dynamics are inherently
connected. Close to the critical point the relaxation time $t_r$ is
related to the correlation length as $t_r \sim \xi^z$, where $z$ is
the dynamical exponent. The random transverse-field Ising spin chain
is very strongly anisotropic at the critical point, since according to
the RG-picture \cite{fisher} and to numerical results \cite{youngrieger}
\be
\ln t_r \sim \xi^{1/2}\;,
\label{scales}
\ee
which corresponds to $z=\infty$. On the other hand the relaxation time
is related to the inverse of the energy-level spacing at the bottom of
the spectrum $t_r \sim (\Delta E)^{-1}$. Then, as a consequence of
eq(\ref{scales}) some energy-like quantities (specific heat, bulk
and surface susceptibilities, etc.) have an essential singularity at
the critical point, while the correlation function of the critical
energy-density has a stretched exponential decay, in contrast to the
usual power law behavior.

Leaving the critical point towards the disordered phase the rare
events with strong correlations still play an important role, until we
reach the region $\delta > \delta_G$, where all transverse-fields are
bigger than the interactions. In the region $0<\delta < \delta_G$,
which is called the Griffiths-McCoy phase the magnetization is a
singular function of the uniform longitudinal field $H_x$ as $m_{\rm
  sing} \sim |H_x|^{1/z}$, where the dynamical exponent $z$ varies
with $\delta$. At the two borders of the Griffiths-McCoy phase it
behaves as $z\approx 1/2\delta\cdot\Bigl(1+{\cal O}(\delta)\Bigr)$
\cite{fisher} as $\delta \to 0$ and $z=1$ as $\delta \to\delta_G^-$,
respectively.

Some of the above mentioned results have been numerically checked
\cite{youngrieger} and in addition various probability distributions
and scaling functions have been numerically determined
\cite{young}. Our present study, which contains analytical and
numerical investigations extends previous work in several respects. In
the case of a limiting random distribution we obtain exact results on
the surface magnetization exponent $x_s=\beta_s/\nu$ and in particular
the thermal exponent $\nu$ through a simple directed walk
consideration. The mapping of the problem of calculating various
universal quantities for the random transverse Ising chain onto the
calculation of statistical properties of appropriately defined random
walks is one of the main achievements of the present paper. With its
help we are also able to explain quantitatively many of the exotic
features of the Griffiths-McCoy region on {\it both} sides of the
transition. 

Moreover, we improved the accuracy of the numerical estimates on the
bulk magnetization exponent in eq(\ref{beta}). Furthermore we present
new results on the magnetization profiles in confined critical systems
as well as about the probability distribution of several quantities.

Throughout the paper we use two types of random distributions.  {\bf
  1)} The binary distribution, in which the couplings can take two
values $\lambda$ and $1/\lambda$ with the probability $p$ and $q=1-p$,
respectively, while the transverse-field is constant:
\beqn
\pi(J)&=&p \delta(J-\lambda)+ q \delta(J-\lambda^{-1})\nonumber\\
\rho(h)&=&\delta(h-h_0)\;,
\label{binary}
\eeqn
The critical point is given by $(p-q) \ln \lambda=\ln h_0$ and
$\delta_G=\sqrt{p/q}$ for $\lambda<1$. {\bf 2)} The uniform
distribution in which the couplings and the fields have uniform
distributions:
\beqn
\pi(J)&=&\cases{1,&for $0<J<1$\cr
                0,&otherwise\cr}\;\nonumber\\
\rho(h)&=&\cases{h_0^{-1},&for $0<h<h_0$\cr
                0,&otherwise\cr}\;.
\label{uniform}
\eeqn
and the critical point is at $h_0=1$ and $\delta_G=\infty$.

The structure of the paper is the following. In Section 2. we present
the free-fermionic description of our model together with the way of
calculation of several physical quantities in this representation. In
Section 3. the surface magnetization and several critical exponents
are calculated exactly using a correspondence with an adsorbing walk
problem.  Phenomenological considerations and numerical estimates
about the distribution of low energy excitations are compared in
Section 4. Numerical results for different critical and off-critical
parameters are presented in Sections 5. and 6., respectively.  Our
results are discussed in the final Section.

\section{Free fermion representation}

We consider the random transverse-field Ising spin chain in
eq(\ref{hamilton}) on a finite chain of length $L$ with free or fixed
boundary conditions, i.e. with $J_L=0$.  The hamiltonian in
eq(\ref{hamilton}) is mapped through a Jordan-Wigner transformation
and a following canonical transformation \cite{liebetal} into a free
fermion model:
\be
H=\sum_{q=1}^L \epsilon_q \left(\eta_q^+\eta_q-{1 \over 2}\right)\;,
\label{fermion}
\ee
where $\eta_q^+$ and $\eta_q$ are fermion creation and annihilation
operators, respectively. The fermion energies $\epsilon_q$ are
obtained via the solution of an eigenvalue problem, which necessitates
the diagonalization of a $2L \times 2L$ tridiagonal matrix {\bf T} with
non-vanishing matrix-elements $T_{2i-1,2i}=T_{2i,2i-1}=h_i$,
$i=1,2,\dots,L$ and $T_{2i,2i+1}= T_{2i+1,2i}=J_i$, $i=1,2,\dots,L-1$.
We denote the components of the eigenvectors $V_q$ as
$V_q(2i-1)=-\phi_q(i)$ and $V_q(2i)=\psi_q(i)$, $i=1,2,\dots,L$, i.e.\
\end{multicols}
\widetext
\noindent\rule{20.5pc}{.1mm}\rule{.1mm}{2mm}\hfill
\be
{\bf T} = \left(
\matrix{
 0  & h_1 &     &       &       &       &     \cr
h_1 &  0  & J_1 &       &       &       &     \cr
    & J_1 &  0  & h_2   &       &       &     \cr
    &     & h_2 &  0    &\ddots &       &     \cr
    &     &      &\ddots&\ddots &J_{L-1}&     \cr
    &     &      &      &J_{L-1}&   0   & h_L \cr
    &     &      &      &       &  h_L  &  0  \cr}
\right)\quad,\qquad
V_q = \left(\matrix{
-\Phi_q(1)\cr
 \Psi_q(1)\cr
-\Phi_q(2)\cr
 \vdots\cr
 \Psi_q(L-1)\cr
-\Phi_q(L)\cr
 \Psi_q(L)}
\right)\quad.
\label{trid}
\ee
\hfill\rule[-2mm]{.1mm}{2mm}\rule{20.5pc}{.1mm}
\begin{multicols}{2} 
\narrowtext
\noindent 
One is confined to the $\epsilon_q \ge 0$ part of the
spectrum\cite{igloiturban96}.

\subsection{Magnetization}

Technically the calculation is more simple, when the boundary condition (b.c.)
does not break the symmetry of the Hamiltonian, therefore we start with a chain
with two free ends. As a consequence, in this case the ground state
expectation value of the local magnetization operator
$\langle0|\sigma_l^x|0\rangle^{\rm free}$ is zero for finite chains.
Then the scaling behavior of the magnetization at the critical point is
obtained from the asymptotic behavior of the (imaginary) time-time
correlation function:
\beqn
G_l(\tau)&=&\langle0|\sigma_l^x(\tau)\sigma_l^x(0)|0\rangle\nonumber\\
&=&\sum_i |\langle i|\sigma_l^x|0\rangle|^2\exp[-\tau(E_i-E_0)]\;,
\label{autocorr}
\eeqn
where $|0\rangle$ and $|i\rangle$ denotes the ground-state and the
$i$-th excited state of $H$ in eq(\ref{fermion}), with energies $E_0$
and $E_i$, respectively. In the thermodynamic limit in the ordered
phase of the system the first excited state is asymptotically
degenerate with the ground-state, thus the sum in eq(\ref{autocorr})
is dominated by the first term. In the large $\tau$ limit
$\lim_{\tau\to\infty} G_l(\tau)=m_l^2$, thus the local magnetization is
given by the off-diagonal matrix-element:
\be
m_l^{\rm free}=\langle1|\sigma_l^x|0\rangle\;.
\label{magnfree}
\ee
In the fermion representation the magnetization operator is expressed as:
\be
\sigma_l^x=A_1 B_1 A_2 B_2 \dots A_{l-1} B_{l-1} A_l\;,
\label{sigmax}
\ee
with
\be
A_i=\sum_q \phi_q(i)(\eta_q^+ + \eta_q)~~~
B_i=\sum_q \psi_q(i)(\eta_q^+ - \eta_q)\;.
\label{ab}
\ee
Using $|1\rangle=\eta_1^+ |0\rangle$ the matrix-element in
eq(\ref{magnfree}) is evaluated by Wick's theorem. Since
for $i \ne j~~\langle0|A_i A_j|0\rangle= \langle0|B_i
B_j|0\rangle=0$ we obtain for the local magnetization
\be
m_l^{\rm free}=\left|\,\matrix{
H_1&G_{11}&G_{12}&\ldots&G_{1l-1}\cr
H_2&G_{21}&G_{22}&\ldots&G_{2l-1}\cr
\vdots&\vdots&\vdots&\ddots&\vdots\cr
H_l&G_{l1}&G_{l2}&\ldots&G_{ll-1}\cr}
\right|\;,
\label{magnfree1}
\ee
where
\beqn
H_j   &=&\langle0|\eta_1 A_j|0\rangle=\Phi_1(j)\nonumber\\
G_{jk}&=&\langle0|B_k A_j|0\rangle=-\sum_{q} \Psi_q(k) \Phi_q(j)\;.
\label{matrelm}
\eeqn
We note that the off-diagonal magnetization $m_l^{\rm free}$ in
eq(\ref{magnfree}) can be used to study the scaling behavior of the
critical magnetization through finite size scaling.

Next we turn to consider the system with symmetry breaking b.c.-s,
when one of the boundary spins is fixed. Then one should formally put
$h_1=0$ or $h_L=0$. For instance if we fix the spin at site $L$ we put
$h_L=0$, implying that $\sigma_L^x$ now commutes with the Hamiltonian
and $S_L^x$ is a good quantum number. In the fermionic
description the two-fold degeneracy of the energy levels,
corresponding to $S_L^x=+1$ and $S_L^x=-1$, is manifested by a
zero energy mode: $\varepsilon_1=0$ in eq(\ref{fermion}), with the
eigenvector $V_1(i)=\delta_{2L,i}$, whereas for $q>1~~V_q(2L)=0$. With
this modification the {\it ground state} expectation value of the
magnetization
\be
m_l^{\rm free+}=\langle0|\sigma_l^x|0\rangle\;,
\label{magnmix}
\ee
is formally given by the determinant in eq(\ref{magnfree1}). The surface
magnetization  $m_s \equiv m_1=\phi_1(1)$ can be computed in a
straightforward manner via the normalization condition $\sum_i
\phi_1^2(i)=1$ leading to the exact formula \cite{peschel}:
\be
m_s=\left[1+\sum_{l=1}^{L-1} \prod_{j=1}^l
\left( h_j \over J_j \right)^2\right]^{-1/2}\;,
\label{peschel}
\ee
where we made use of the fact that $\varepsilon_1=0$ because of the
exact degeneracy of the ground state.

Next we consider those boundary conditions, when both boundary spins
are fixed. Then one should formally put $h_1=0$ and $h_L=0$.  This
situation is, however, more complicated than the mixed b.c., since it
describes both parallel ($++$) and anti-parallel ($+-$) boundary
conditions. To determine the magnetization profiles in these cases we
make use the duality properties of the quantum Ising model. First we
define the dual Pauli-operators $\tau_{i+1/2}^x,\tau_{i+1/2}^z$ as:
\be
\tau_{i+1/2}^z=\sigma_i^x \sigma_{i+1}^x~~,~~\sigma_i^z=
\tau_{i-1/2}^x\tau_{i+1/2}^x\;,
\label{dualpauli}
\ee
in terms of which the Hamiltonian in eq(\ref{hamilton}) is expressed as:
\be
H=-\sum_l J_l\tau_{l+1/2}^z -\sum_l h_l\tau_{l-1/2}^x\tau_{l+1/2}^x\;.
\label{dualhamilton}
\ee
In the dual model formally the couplings and fields are interchanged,
thus the dual Hamiltonian has zero surface fields, since $J_0=J_L=0$
in eq(\ref{hamilton}). Then it can be easily shown that the even
sector (i.e. those states which contain even number of fermions) of
the free chain corresponds to the (++) parallel boundary condition of
the dual model, whereas the odd sector of the free chain to the (+-)
anti-parallel boundary condition.

To obtain the magnetization profile for fixed boundary spin conditions
we use the expression
\be
\tau_{l+1/2}^x=\sigma_0^z\sigma_1^z\dots\sigma_l^z\;,
\label{dualx}
\ee
which can be obtained from eq(\ref{dualpauli}), then express the
product of $\sigma^z$-s by fermion operators through the relation
$\sigma_i^z=A_i B_i$ and evaluate the r.h.s. of eq(\ref{dualx}) in the
corresponding free-chain situation. For the parallel spin boundary
condition we take the free-chain vacuum expectation value:
\be
m_l^{++}=\langle0|\tau_{l+1/2}^x|0\rangle^{++}
=\langle0|A_0 B_0 A_1 B_1 \dots A_l B_l|0\rangle^{\rm free}\;,
\label{magn++}
\ee
which can be expressed through Wick's theorem as:
\be
m_l^{++}=\left|\,\matrix{
1&0&0&\ldots&0\cr
0&\tilde{G}_{11}&\tilde{G}_{12}&\ldots&\tilde{G}_{1l}\cr
\vdots&\vdots&\vdots&\ddots&\vdots\cr
0&\tilde{G}_{l1}&\tilde{G}_{l2}&\ldots&\tilde{G}_{ll}\cr}
\right|\;.
\label{magn++1}
\ee
Here $\tilde{G}_{jk}$ is the same as in eq(\ref{matrelm}), however it
is calculated with the dual couplings $h_l \leftrightarrow J_l$.

To obtain the magnetization profile in the anti-parallel spin boundary
condition (+-) one should take the expectation value of the
r.h.s. of eq(\ref{dualx}) in the lowest state of the odd sector of the
free chain, which is the first excited state of the Hamiltonian:
$|1\rangle=\eta_1^{+}|0\rangle$. Thus:
\be
m_l^{+-}=\langle0|\tau_{l+1/2}^x|0\rangle^{+-}
=\langle1|A_0 B_0 A_1 B_1 \dots A_l B_l|1\rangle^{\rm free}\;.
\label{magn+-}
\ee
To evaluate this expression by Wick's theorem one should notice that
the state $|1\rangle=\eta_1^{+}|0\rangle$ can be considered the vacuum state of
such a system, where $\eta_1$ and $\eta_1^{+}$ are interchanged, which
formally means that $\psi_1(k) \to -\psi_1(k)$ and $\epsilon_1 \to
-\epsilon_1$. Then the expectation values in eq(\ref{matrelm}) are
modified as
\be
\overline{G}_{jk}=\langle1|B_k A_j|1\rangle=
-\sum_{q>1} \Psi_q(k) \Phi_q(j)+\Psi_1(k) \Phi_1(j)\;,
\label{matrelm+-}
\ee
which again have to be evaluated with the dual couplings $h_l
\leftrightarrow J_l$. Then the $m_l^{+-}$ profile is given by the
determinant in eq(\ref{magn++1}), where the matrix-elements
$\tilde{G}_{jk}$ are replaced by $\overline{G}_{jk}$.

\subsection{Susceptibility and Autocorrelations}

The local susceptibility $\chi_l$ at site $l$
is defined through the local magnetization $m_l$ as:
\be
\chi_l=\lim_{H_l \to 0} {\delta m_l \over \delta H_l}\;,
\label{locsusc}
\ee
where $H_l$ is the strength of the local longitudinal field, which
enters in the Hamiltonian in eq(\ref{hamilton}) as $H_l
\sigma_l^x$. $\chi_l$ can be expressed as:
\be
\chi_l=2 \sum_i{|\langle i|\sigma_l^x|0\rangle|^2\over E_i-E_0}\;,
\label{locsusc1}
\ee
which for boundary spins is simply given by:
\be
\chi_{1}=2 \sum_q {|\phi_q(1)|^2 \over \epsilon_q}\;.
\label{surfsusc}
\ee

Next we consider the dynamical correlations of the system as a function
of the imaginary time $\tau$. First,
we note that the correlations between surface spins can be obtained
directly from eq(\ref{autocorr}) as:
\be
G_1(\tau)=\sum_q | \Phi_q(1)|^2 \exp(-\tau \epsilon_q)\;.
\label{surfcorr}
\ee
For bulk spins the matrix-element $\langle i|\sigma_l^x|0\rangle$ in
eq(\ref{autocorr}) is more complicated to evaluate, therefore one goes
back to the first equation of (\ref{autocorr}) and considers the
time-evolution in the Heisenberg picture:
\beqn
\sigma_l^x(\tau)&=&\exp(\tau H) \sigma_l^x \exp(-\tau H)\nonumber\\
&=&A_1(\tau) B_1(\tau)
\dots A_{l-1}(\tau) B_{l-1}(\tau) A_l(\tau)\;.
\label{sigmatau}
\eeqn

The general time and position dependent correlation function
\be
\langle\sigma_l^x(\tau) \sigma_{l+n}^x\rangle
=\langle A_1(\tau) B_1(\tau) \cdots A_l(\tau) A_1 B_1\dots A_{l+n}\rangle\;,
\label{gencorr}
\ee
can then be evaluated by Wick's theorem as a product of two-operator
expectation values, which in turn is written into the compact form
as a Pfaffian:
\end{multicols}
\widetext
\noindent\rule{20.5pc}{.1mm}\rule{.1mm}{2mm}\hfill
\beqn
\langle\sigma_l^x(\tau) \sigma_{l+n}^x\rangle
&=&\left.
\matrix{|\;
\langle A_1(\tau)B_1(\tau)\rangle &
\langle A_1(\tau)A_2(\tau)\rangle &
\langle A_1(\tau)B_2(\tau)\rangle &
\cdots\quad
\langle A_1(\tau)A_l(\tau)\rangle &
\langle A_1(\tau)A_1\rangle &
\cdots\quad
\langle A_1(\tau)A_{l+n}\rangle\cr
&
\langle B_1(\tau)A_2(\tau)\rangle &
\langle B_1(\tau)B_2(\tau)\rangle &
\cdots\quad
\langle B_1(\tau)A_l(\tau)\rangle &
\langle B_1(\tau)A_1\rangle &
\cdots\quad
\langle B_1(\tau)A_{l+n}\rangle\cr
& &
\langle A_2(\tau)B_2(\tau)\rangle &
\cdots\quad
\langle A_2(\tau)A_l(\tau)\rangle &
\langle A_2(\tau)A_1\rangle &
\cdots\quad
\langle A_2(\tau)A_{l+n}\rangle\cr
& &         & \ddots & & \vdots\cr
& &         & & &  \langle B_{l+n-1}A_{l+n}\rangle}
\right\vert\nonumber\\
&=&
\pm \left[ {\rm det}\, C_{ij}\right]^{1/2}\;,
\label{pfaffian}
\eeqn
\hfill\rule[-2mm]{.1mm}{2mm}\rule{20.5pc}{.1mm}
\begin{multicols}{2} 
\narrowtext
\noindent 
where $C_{ij}$ is an antisymmetric matrix $C_{ij}=-C_{ji}$, with the
elements of the Pfaffian (\ref{pfaffian}) above the diagonal.
%
%
%
%
At zero temperature the elements of the
Pfaffian are the following:
\begin{eqnarray}
\langle A_j(\tau) A_k\rangle
& = &\sum_q \Phi_q(j) \Phi_q(k) \exp(-\tau \epsilon_q)\;,
\nonumber\\
\langle A_j(\tau) B_k\rangle
& = &\sum_q \Phi_q(j) \Psi_q(k) \exp(-\tau \epsilon_q)\;,
\nonumber\\
\langle B_j(\tau) B_k\rangle
& = &-\sum_q \Psi_q(j) \Psi_q(k) \exp(-\tau \epsilon_q)\;,
\nonumber\\
\langle B_j(\tau) A_k\rangle
& = &-\sum_q \Psi_q(j) \Phi_q(k) \exp(-\tau \epsilon_q)\;,
\label{pfaffelm}
\end{eqnarray}
whereas the equal-time contractions are given in (\ref{ab}). For the
finite temperature contractions see c.f. \cite{stolze}.

\section{Surface magnetization and the mapping to adsorbing walks}

In this Section we analyze the surface magnetization of the RTIM
using a mapping to an adsorbing random walk problem. In this way we obtain
exact results for the critical exponents $\beta_s$ and $\nu$, as well as
for the surface magnetization scaling dimension $x_m^s$.
These observations will then be used in the following Section to identify
the structure of the strongly coupled domains (SCD), which are responsible
for the low energy excitations in the system.

\subsection{Surface magnetization and correlation length}

The surface magnetization in eq(\ref{peschel}) represents perhaps the
simplest order-parameter of the transverse-field Ising chain. Note that
the scaling behavior of end-to-end correlations
\be
C_L=[\langle\sigma_1^x\sigma_L^x\rangle]_{\rm av}
\ee
is identical to that of the surface magnetization since $C_L$ as well
as $m_s$ involve only surface spin operators that have anomalous
dimension $x_s$. So for instance $C_L(\delta=0)\sim L^{-2x_s}$ and
$C_{L\to\infty}(\delta)\sim\delta^{2\beta_s}$.

\begin{figure}
\epsfxsize=\columnwidth\epsfbox{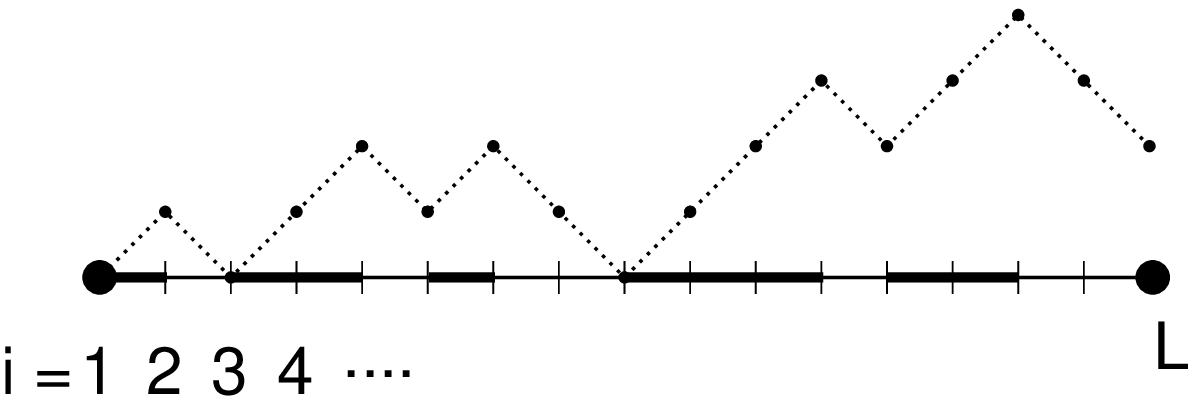}
\caption{Sketch of the correspondence between a bond configuration and
  a random walk. The thick segments in the horizontal line indicate
  strong (\protect{$J_i=\lambda^{-1}$}) bonds, thin segments weak
  bonds (\protect{$J_i=\lambda$}). This bond configuration corresponds
  to a surviving walk (as indicated by the broken line staying
  completely above the horizontal line), implying a finite surface
  magnetization. Actually for this example \protect{$m_s=1/3^{1/2}$} in
    the case $\lambda=0$, since the random walk touches the horizontal
    line three times (including the starting point).}
\label{fig1}
\end{figure}

In the following we determine its average behavior for the symmetric
binary distribution, i.e. with p=q=1/2 in eq(\ref{binary}).  First we
consider the system of a large, but finite length $L$ at the critical
point. In the limit $\lambda \to 0$ the surface magnetization in
eq(\ref{peschel}), which is expressed as a sum of products has a
simple structure. Considering a random realization of the couplings
the surface magnetization is zero, whenever a product of the form of
$\prod_{i=1}^l J_i^{-2}$, $l=1,2,\dots,L$ is infinite, i.e. the number
of $\lambda$ couplings exceeds the number of $\lambda^{-1}$ couplings
in any of the $[1,l]$ intervals. On the other hand, if the surface
magnetization has a finite value, it could be of the form
$m_s=1/(n+1)^{1/2}$, where $n=0,1,2,\dots$ measures the number of
intervals, which are characterized by having the same number of
$\lambda$ and $\lambda^{-1}$ couplings. To have a more transparent
picture we represent the distribution of couplings by directed walks,
which start at zero and make the $i$-th steps upwards (for a coupling
$J_i=\lambda^{-1}$) or downwards (for a coupling $J_i=\lambda$). As
illustrated in Fig.1 the surface magnetization corresponding to a
random sequence is non-zero, only if the representing walk does not go
below the `time'-axis, while the corresponding surface magnetization
is given as $m_s=1/(n+1)^{1/2}$, where now $n$ just counts the number
how many times the walk has touched the $t$-axis for $t>0$.

Then the ratio of walks representing a sample with finite surface
magnetization is just the survival probability of the walk $P_{\rm surv}$,
which is given by $P_{\rm surv}(L)\propto L^{-1/2}$, for walks (samples)
of length $L$. In the thermodynamic limit this probability vanishes,
thus the $typical$ realization of the chain, i.e. the event with
probability one has zero surface magnetization. This is certainly
different from the $average$ value, which is dominated by the rare
events represented by surviving walks with an $m_s=O(1)$.
Consequently the average surface magnetization of a critical chain of
length $L$ is given by
\be
[m_s(L,\delta=0)]_{av}=A L^{-1/2}+O(L^{-3/2})\;.
\label{smagnavr}
\ee
On the other hand one knows from finite size scaling that
$[m_s(L,\delta=0)]_{av} \sim L^{-x_m^s}$, where $x_m^s=\beta_s/\nu$ is
the scaling dimension of the surface magnetization. Thus from
eq(\ref{smagnavr}) we can read the {\it exact result}:
\be
x_m^s={1 \over 2}\;,
\label{xms}
\ee
while the prefactor in eq(\ref{smagnavr}) is obtained as $A=.6431961$
from a numerical calculation.

In the following we argue that the exponent in eq(\ref{xms}) is the
same far all values $\lambda$ of the binary distribution. First we
mention that the finite-size critical surface magnetization
$[m_s^{\lambda}(L,0)]_{av}$ is a monotonically decreasing function of
$\lambda \le 1$: $[m_s^{\lambda_1}(L,0)]_{av}<
[m_s^{\lambda_2}(L,0)]_{av}~~0 \le \lambda_1 < \lambda_2 \le 1$ for
any value of $L$. Thus the corresponding exponents also satisfy
$x_m^s(\lambda_1) \le x_m^s(\lambda_2)$. However according to exact
results the value of $x_m^s$ is the same, i.e. $1/2$ for the
homogeneous model \cite{surfising} $(\lambda=1)$ and for the extreme
inhomogeneous model $(\lambda \to 0)$, thus the relation in
eq(\ref{xms}) should hold for any parameters of the distribution.  We
note that this is the first exact derivation of the $x_m^s$ exponent,
the value of which is in agreement with previously known exact and
conjectured results in eqs(\ref{betas}) and (\ref{nu}).

Next we calculate the $\nu$ exponent from the $\delta$ dependence of the
surface magnetization. In the scaling limit $L \gg 1$, $|\delta| \ll 1$
the surface magnetization can be written as:
\be
[m_s^{\lambda}(L,\delta)]_{av}=[m_s^{\lambda}(L,0)]_{av}
\tilde{m}_s(\delta L^{1/\nu})\;.
\label{msscale}
\ee
Here the scaling function $\tilde{m}_s(y)$, which depends on the ratio
of $L$ and the correlation length $\xi$, can be expanded as
$\tilde{m}_s(y)=1+By+O(y^2)$, so that one obtains for the $\delta$
correction to the surface magnetization:
\be
[m_s(L,\delta)]_{av}-[m_s(L,0)]_{av} \propto
\delta L^{\Theta}\;,
\label{deltacorr}
\ee
with $\Theta=1/\nu-x_m^s$. This exponent can also be determined in the
$\lambda \to 0$ limit of the binary distribution. Now, slightly
outside the critical point the products of $l$ terms in the sum of
eq(\ref{peschel}) will contain a factor of $(1+\delta)^{2l}\simeq
1+2l\delta$ in leading order of $\delta$. Then the surface
magnetization of a coupling distribution which is represented by a
surviving walk is given by:
\beqn
m_s&=&\left[1+\sum_{i=1}^n(1+2l_i \delta)\right]^{-1/2}\nonumber\\
&=&(1+n)^{-1/2}-\delta {\sum_i l_i \over (n+1)^{3/2}}+O(\delta^2)\;,
\label{msdelta}
\eeqn
where $l_i$ gives the position of the $i$-th touching point of the
walk with the $t$-axis. Next we consider a typical surviving walk,
which has $n=O(1)$ return points,(since the probability of $n$ returns
decreases exponentially) and these points are situated at
$l_i=O(L^{1/2})$. Consequently for a typical surviving walk the
correction term in eq(\ref{msdelta}) is $O(L^{1/2})$, what should be
multiplied by the surviving probability of $O(L^{-1/2})$. Since the
surviving walks have a sharp probability distribution we are left with
the result: $[m_s(L,\delta)]_{av}-[m_s^{\lambda}(L,0)]_{av}= B
\delta+O(\delta^2)$, where the constant is given from numerical
calculations as: $B=0.270563 \simeq 17/20 \pi$. Comparing our result
with that in eq(\ref{deltacorr}) we get for the correlation length
critical exponent 
\be
\nu=2\;.
\ee
This is the first exact determination of the $\nu$ exponent, previously
conjectured in Fisher's RG-study\cite{fisher}.

\subsection{Relation between surface magnetization and adsorbing random walks}

Here we summarize and extend the mapping between the surface
magnetization of the RTIM and the surviving probability of the
corresponding adsorbing random walk. First, we consider again the
extreme binary distribution of couplings in eq(\ref{binary}) with
$h_0=1$, such that the control parameter in eq(\ref{delta}) is given
by:
\be
\delta={q-p \over 4pq } {1 \over \ln \lambda}\;.
\label{deltabin}
\ee
Then, according to the considerations of the previous section, the
corresponding adsorbing walk has an asymmetric character: it makes
steps with probabilities $p$ and $q=1-p$ off and towards the wall,
respectively. The corresponding control parameter $\delta_w=q-p$
in eq(\ref{deltaw}) is proportional to $\delta$ in eq(\ref{deltabin}).
Our basic observation can be summarized as:
\be
m_s(\delta,L) \sim P_{\rm surv}(\delta_w,L),~~~\delta \sim \delta_w\;.
\label{relation}
\ee

{\it At the critical point} from eq(\ref{psurv0}) $P_{\rm surv}(0,L) \sim
L^{-\gamma}$ and $\gamma=1/2$ is just $x_m^s$, the surface
magnetization scaling dimension of the RTIM.  {\it In the paramagnetic
phase} of the RTIM $\delta>0$ and the corresponding walk has a
drift towards the adsorbing wall. The surviving probability in
eq(\ref{psurv+})
\beqn
P_{\rm surv}(\delta_w>0,L) & \sim & \exp(-L/\xi_w),~~~\nonumber\\
\xi_w&=&{8pq \over (p-q)^2}
\sim\delta_w^{-2}\;,
\label{psurvp}
\eeqn
is characterized by a correlation length, which diverges as
$\xi_w\sim\delta^{-\nu}$
with $\nu=2$. We note that the expression for $\xi_w$ in (\ref{psurvp})
agrees with the RTIM result by Fisher \cite{fisher}.
Finally, {\it in the ferromagnetic phase} $\delta<0$ the corresponding
walk drifts off the wall and the surviving probability has a
finite limit as $L \to \infty$:
\be
P_{\rm surv}(\delta<0,L \to \infty)={p-q \over p} \sim -\delta_w\;.
\label{psurvf}
\ee
This expression then corresponds to a finite average surface magnetization
of the RTIM, which linearly vanishes at the critical point. Thus the
surface magnetization exponent
of the RTIM is $\beta_s=1$, in agreement with the exact results by McCoy
and Wu \cite{mccoywu}.

These results obtained for the extreme binary distribution can be
generalized for other random distributions, too. It is enough to
notice, that the surface magnetization in a sample with non-surviving
walk character is exponentially vanishing with the size
of the system. Therefore the basic relation in eq(\ref{relation})
remains valid.  Then at the critical point the $P_{\rm
surv}(\delta=0,L) \sim L^{-1/2}$ is a consequence of the Gaussian
nature of the random walk. Similarly, the relations $\xi_w \sim
\delta_w^{-2},~~\delta_w>0$ and $P_{\rm surv}(\delta_w<0,L) \sim -
\delta_w,~~\delta_w<0$ follow from the scaling behavior of
the random walks.

\section{Distribution of the low-energy excitations}

In the previous Section we saw that the surface order is connected to
such a coupling distribution, which can be represented by a surviving
walk.  Since (local) order and small (vanishing) excitation energies
are always connected, we can thus identify the local distribution of
couplings, which result in a strongly coupled domain (SCD). Note that
a SCD is not simply a domain of strong bonds, but it generally has a
much larger spatial extent.

To estimate the excitation energy $\epsilon$ of an SCD we make use the
exact result for the lowest gap $\epsilon_1(l)$ of an Ising quantum
chain of $l$ spins with free b.c. \cite{iktsz}: Since we are interested
in a bond- and field configuration that gives rise to an exponentially
small gap $\epsilon_1$ we can neglect the r.h.s.\ of the eigenvalue
equation \be {\bf T}\cdot V_1 = \epsilon_1 V_1\;, \ee c.f.\ 
eq(\ref{trid}) and derive approximate expressions for the
eigenfunctions $\Phi_1$ and $\Psi_1$. With these one arrives at
\be 
\epsilon_1(l) \sim m_s \overline{m_s} 
\prod_{i=1}^{l-1} {h_i\over J_i}\;.
\label{lambda1}
\ee
Here $m_s$ and $\overline{m_s}$ denote the finite-size surface
magnetizations at both ends of the chain, as defined in
eq(\ref{peschel}) (for $\overline{m_s}$ simply replace $h_j/J_j$ by
$h_{L-j}/J_{L-j}$ in this eq.). If the sample has a low-energy
excitation, then both ends surface magnetizations are of $O(1)$,
consequently the coupling distribution follows a surviving walk
picture. The corresponding gap estimated from eq(\ref{lambda1}) is
given by:
\be
\epsilon_1 \sim \prod_{i=1}^{l-1} {h_i \over J_i} \sim
\exp\Bigr\{-l_{\rm tr}\cdot\overline{\ln(J/h)}\Bigl\}\;,
\label{gapestimate}
\ee
where $l_{\rm tr}$ measures the size of transverse fluctuations of a
surviving walk of length $l$ and $\overline{\ln(J/h)}$ is an average
coupling.  In the following we assume that the excitation energy of
surface SCD-s are of the same order of magnitude as those localized in
the bulk of the system and have the same type of coupling
distributions.  Thus we identify the SCD-s, both at the surface and in
the volume of the system, as a realization of couplings and fields with
surviving walk character and having an excitation energy given
eq(\ref{gapestimate}).  With this prerequisite we are now ready to
apply our theory for the critical and off-critical regions of the RTIM
of $L$ sites.

At the bulk critical point the characteristic length $l$ of surviving
regions is of the order of the size of the system $L$, thus the
SCD extends over the volume of the system. The transverse fluctuations
of the couplings in the SCD are from eq(\ref{trfluct0}) as $l_{\rm tr}
\sim L^{1/2}$, thus we obtain for the scaling relation of the energy
gap at the critical point:
\be
\epsilon(\delta=0,L) \sim \exp(- {\rm const}\cdot L^{1/2})\;,
\label{epscrit}
\ee
in accordance with the existing numerical results\cite{youngrieger}.
At this point it is useful to point out the origin of the exponent
$1/2$ accompanying the length scale $L$ in eq(\ref{epscrit}): it is
the fact that the sequence of $h_i/J_i$ is random and uncorrelated,
for a general sequence one would have $l_{\rm tr}\sim L^{\omega}$ with
$\omega$ being the wandering exponent (the scaling dimension of the
transverse fluctuations) of the particular sequence under
considerations. For instance one could also consider relevant
aperiodic sequence (generated in a deterministic fashion), which have
either the same or different wandering exponents, leading either to
the same or a different scaling behavior of the energy scale at the
critical point as the random chain studied here \cite{aperiodic}.

\begin{figure}[t]
\epsfxsize=\columnwidth\epsfbox{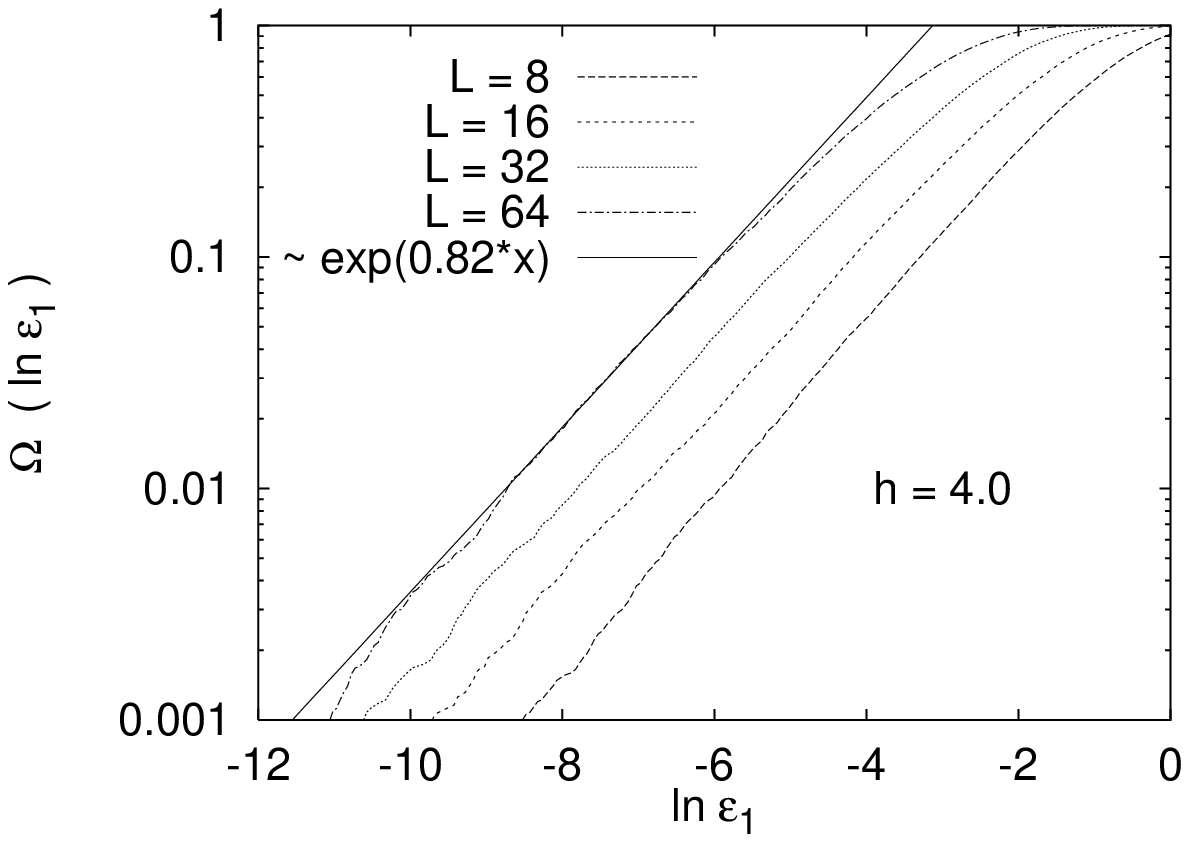}
\epsfxsize=\columnwidth\epsfbox{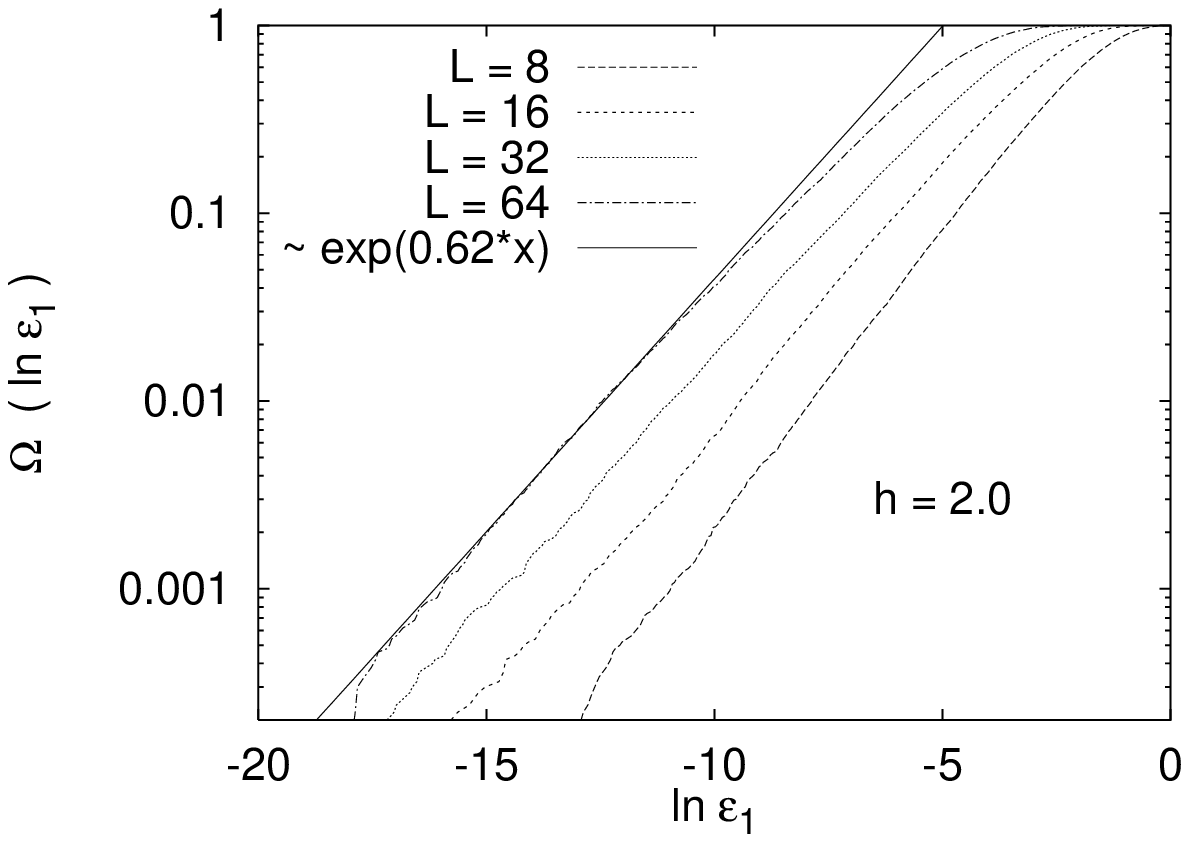}
\epsfxsize=\columnwidth\epsfbox{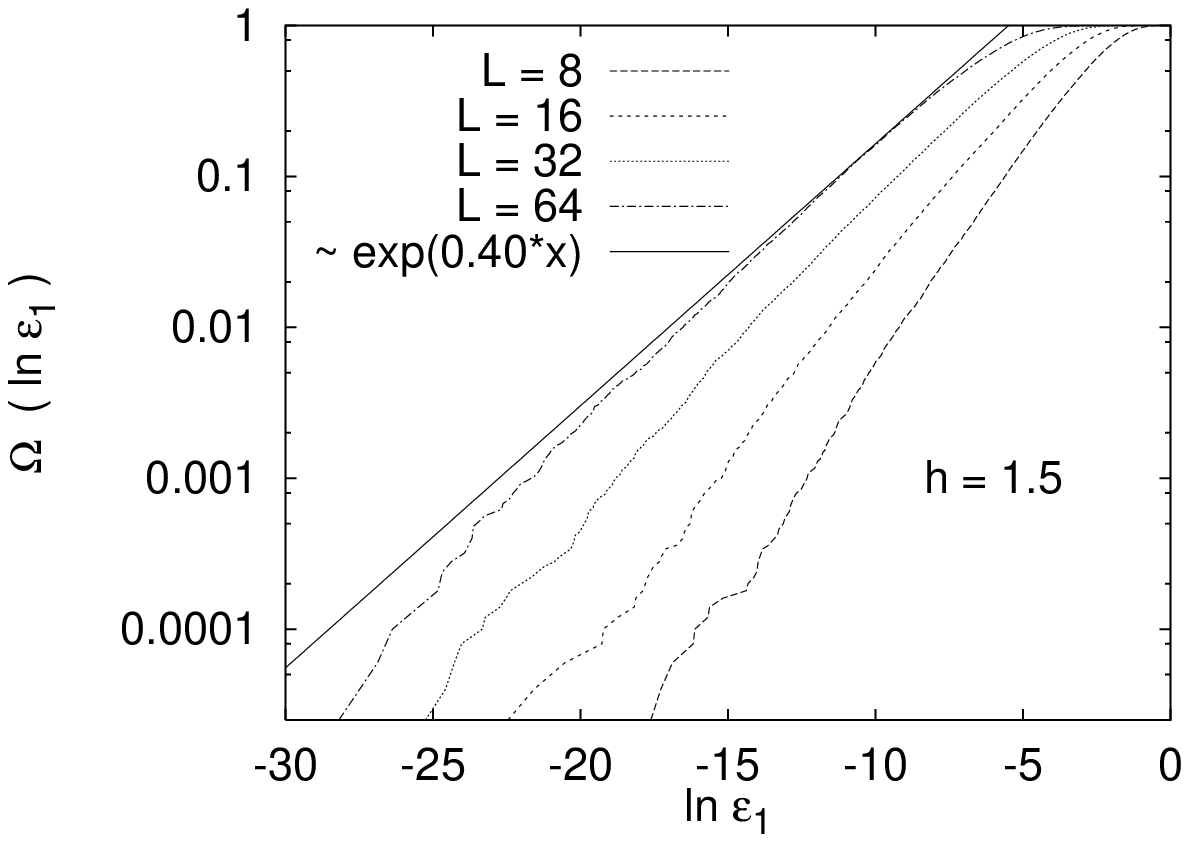}
\caption{ The integrated gap probability distribution $\Omega_L(\ln
\epsilon_1)$ in the disordered phase ($h>1$) for different values of
$h$.  The dynamical exponent $z(h)$ is extracted from the expected
asymptotic form $\ln\Omega_L(\ln\epsilon_1) = 1/z(h) \ln\epsilon_1) +
{\rm const}$ which is a straight line when using a logarithmic scale
on the y-axes.  Thus $1/z(h)\approx0.82$, $0.62$ and $0.40$ for
$h=4.0$, $2.0$ and $1.5$, respectively. The data for the uniform
distribution averaged over 50000 samples. Note that for large $h$,
i.e. far away from the critical point, there is essentially no system
size dependence, whereas closer to the critical point the asymptotic
slope can only is reached only for large enough system sizes.}
\label{fig2}
\end{figure}

In the paramagnetic phase the probability to find a SCD of size $l$,
which is localized at a given point is proportional to $\exp(-l/\xi)$,
c.f.\ eq(\ref{psurvp}).  Since the SCD can be located at any point
of the chain, the actual probability is proportional to the length of
the chain, thus $P_L(l) \sim L \exp(-l/\xi)$. The characteristic size
of SCD-s obtained from the condition $P_L(l)=O(1)$ is given by:
\be
l \sim \xi \ln L~,~~\delta>0\;,
\label{lpara}
\ee
which grows very slowly with the linear size of the system. The
characteristic transverse fluctuations of such a walk is given -
according to eq(\ref{trfluctmax}) - as $l_{\rm tr} \approx (q-p) l
\alpha$, with $\alpha=O(1)$. Putting this expression into
eq(\ref{gapestimate}) we obtain the scaling relation
\be
\epsilon(\delta>0,L) \sim L^{-z(\delta)}\;,
\label{epspara}
\ee
where the dynamical exponent $z(\delta)=2 \alpha/\delta$ is a
continuous function of the control parameter $\delta$. Our estimate
qualitatively agrees with Fisher's result\cite{fisher}, that close to
the critical point $z(\delta)=1/2\delta$.

The dynamical exponent $z(\delta)$ is conveniently measured from the
scaling behavior of the probability distribution $P_L(l) \sim
P_L(\ln(\epsilon)) \sim L$. For a given large $L$ the scaling
combination from eq(\ref{epspara}) is $L \epsilon^{1/z}$, thus
\be
P(\epsilon) \sim \epsilon^{-1+1/z(\delta)}\;.
\label{gapdistr}
\ee
The distribution function of the gap in eq(\ref{gapdistr}) has already been
studied in\cite{youngrieger} for periodic boundary conditions. Here we
considered free chains and investigated the accumulated probability
distribution
\be
\Omega_L(\ln \epsilon) = \int_{-\infty}^{\ln \epsilon} dy P_L(y)\;.
\label{accum}
\ee
As seen on Fig. \ref{fig2} the accumulated probability distribution for low
energies is approximately a straight line on a log-log plot and from
the slope one can estimate $1/z(\delta)$ quite accurately.

\begin{figure}[t]
\epsfxsize=\columnwidth\epsfbox{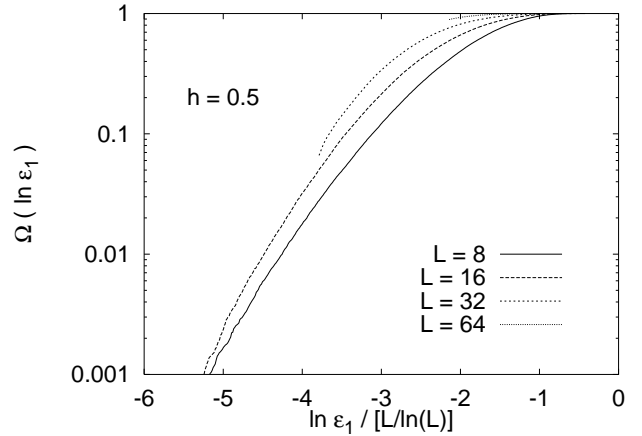}
\caption{The same as in the Fig. 2 in the {\it ordered} phase
($h<1$). The data do not scale with $\ln \epsilon_1/L$, there are
strong logarithmic corrections. A scaling with $\ln
\epsilon_1/[L/ln(L)]$ is also poor (as can be seen in the figure),
most probably higher powers of $\ln(L)$ are involved.}
\label{fig3}
\end{figure}

In the ferromagnetic phase of the RTIM the size of the SCD is of the
order of the sample, $l \sim L$ and also the transverse fluctuations
of the couplings are $l_{\rm tr} \sim L$. Consequently the energy of the
first excitations scale exponentially with the size of the system:
$\epsilon \sim \exp(-\rm{const} L)$. Here, however, one should
take into account that - due to the duality relation in
eq(\ref{dualhamilton}) - in a strongly coupled environment there are
always weakly coupled domains (WCD), which are the counterparts of the
SCD in the paramagnetic phase. The characteristic size of a WCD is
$O(\ln L)$, and their presence will reduce the size of the SCD-s, such
that one expects logarithmic corrections to the size of transverse
fluctuations $l_{\rm tr}$. Indeed the numerical results on the accumulated
gap distribution function in Fig. \ref{fig3} can be interpreted with the
presence of such corrections.

\begin{figure}[t]
\epsfxsize=\columnwidth\epsfbox{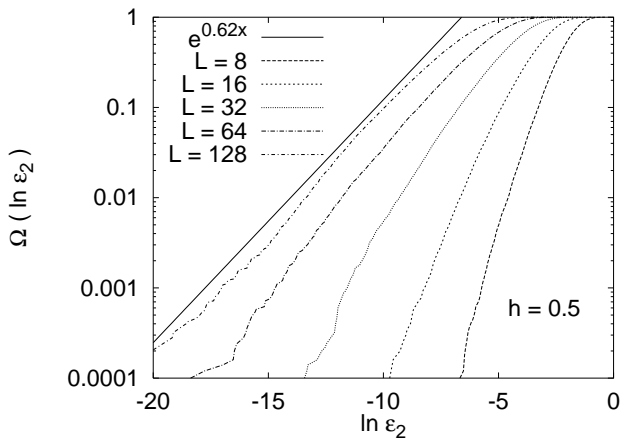}
\caption{The same as in Fig. 3 for the second lowest excitation, i.e.\
$\Omega_L(\ln \epsilon_2)$. One observes that asymptotically
$\ln\Omega_L(\ln\epsilon_2)=1/z(h) \ln\epsilon_2 + {\rm const}$, with
$z(h=0.5)=0.62=z(h=2.0)$ as one would expect from duality, by which
$z(h)=z(1/h)$.}
\label{fig4}
\end{figure}

In the ferromagnetic phase many physical quantities (connected
autocorrelation function, susceptibility, etc.) are connected with the
distribution of the second gap. Unfortunately, we can not make an
estimate for $\epsilon_2$ on the base of our present approach. However,
our model is self-dual, the distributions of the couplings and the
fields transform to each other in eq(\ref{dualhamilton}) for
$\delta \to - \delta$. Therefore,
we assume that the scaling behavior of $\epsilon_1$ in the paramagnetic
phase and that of $\epsilon_2$ in the ferromagnetic phase are also
related through duality, thus
\be
\epsilon_2(\delta<0,L) \sim L^{-z(-\delta)}\;.
\label{lambda2}
\ee
Indeed, as seen on Fig. \ref{fig4} the scaling relation in
eq(\ref{lambda2}) is satisfied, however with strong, logarithmic
corrections.
\vspace{1cm}

\section{Critical properties}

\subsection{Surface magnetization - canonical vs. micro-canonical ensemble}

The surface magnetization of the RTIM has already been studied in
Section 3.  Here we revisit this problem in order to answer to the
question, whether the values of the average quantities and the
corresponding critical exponents depend or not on the ensemble used in
the calculations. Our present study is motivated by a recent work
\cite{pazmandi} in which finite-size scaling methods and their
predictions for critical exponents \cite{chayes} have been scrutinized
for random systems.

In our approach in Section 3 the bond-- and field configurations were
taken completely random according to the corresponding distribution.
We call this the {\it canonical} ensemble, since only the ensemble
average of $\ln J_i$ and $\ln h_i$ is held fixed. One can also confine
oneself on a subset of this ensemble, in which the the product of all
bonds in the chain is exactly equal to the product of all fields in
the chain, i.e.\ $\sum_{i=1}^{L-1}\ln J_i = \sum_{i=1}^{L-1}\ln h_i$
(note that we study the surface magnetization in a chain with one
fixed boundary condition such that there are exactly as many bonds as
fields). This we call the {\it micro-canonical} ensemble. The
motivation for the introduction of this ensemble can be found in
\cite{pazmandi}: essentially it is a more restrictive way of
fulfilling the criterium $[\ln J]_{\rm av}=[\ln h]_{\rm av}$ for being
at the critical point.

The critical exponents of the canonical ensemble
eqs(\ref{xms}),(\ref{nu}), which agree with other exact and RG
results, are then the canonical ones.

\begin{figure}
\epsfxsize=\columnwidth\epsfbox{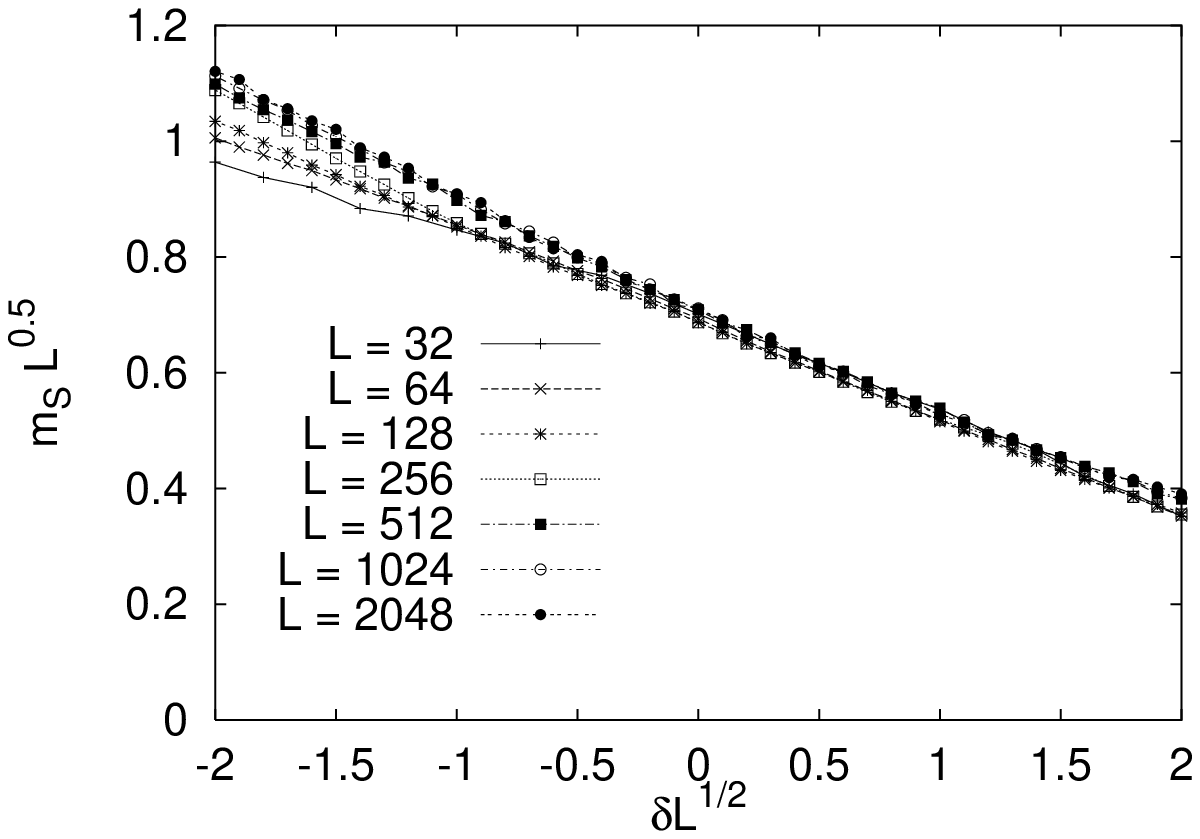}
\epsfxsize=\columnwidth\epsfbox{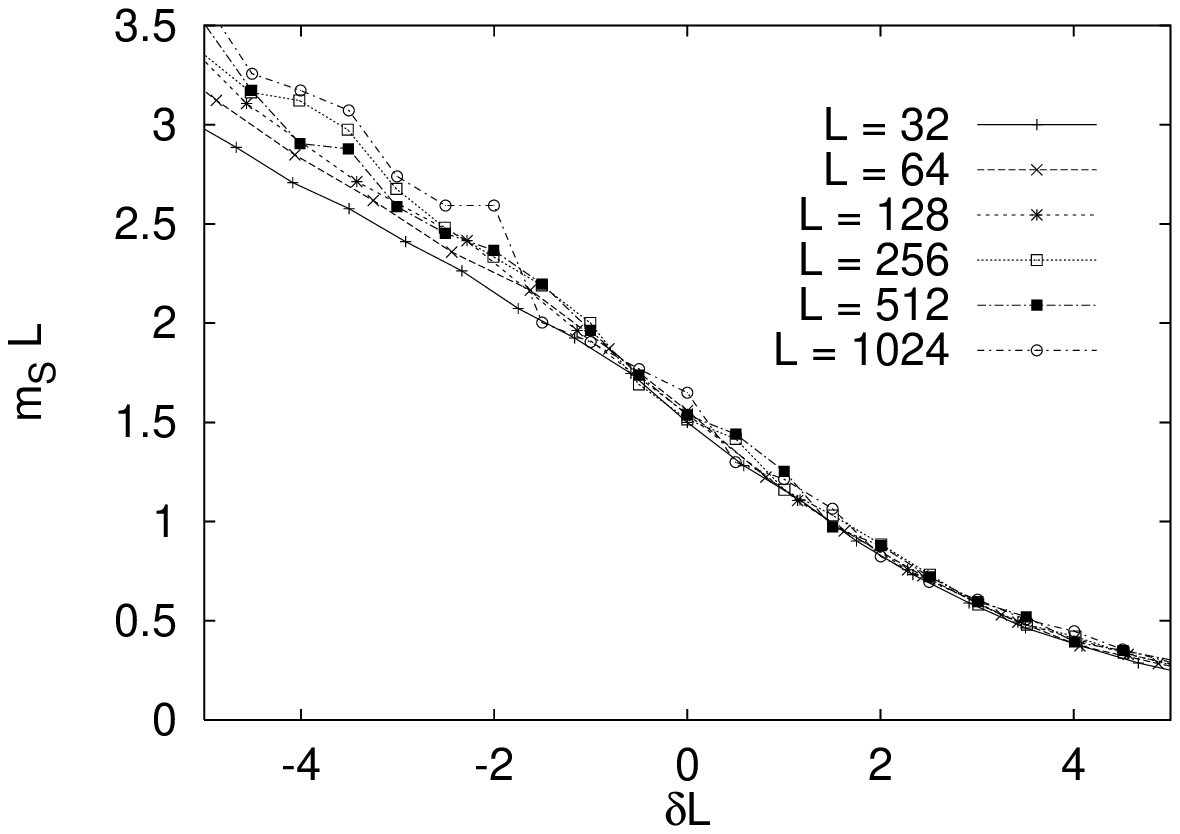}
\caption{Scaling plot of the surface magnetization in the canonical
ensemble (top) and the micro-canonical (bottom) ensemble. Both data
are for the uniform distribution averaged over 100000 samples.}
\label{fig5}
\end{figure}

In the following we calculate the critical exponents corresponding to
eqs(\ref{xms}),(\ref{nu}) also in the micro-canonical ensemble. We use
again the symmetric binary distribution in eq(\ref{binary}), such that
the samples have the same number of $\lambda$ and $\lambda^{-1}$
couplings. In the extreme limit $\lambda \to 0$, as in the
canonical case, the critical point surface magnetization can be
determined exactly, through studying the surviving probability of the
corresponding adsorbing random walk.  To determine the micro-canonical
surviving probability first we note that from the canonical walks only
a fraction of $O(L^{-1/2})$ is micro-canonical. Second, the
micro-canonical surviving walks have their end at the starting
point. Such returning surviving walks are of a fraction of
$O(L^{-3/2})$ among the canonical walks. Thus the surviving
probability of micro-canonical walks is $P_{\rm surv}({\rm m.c.}) \sim L^{-1}$,
therefore the micro-canonical surface magnetization scaling dimension
is given by:
\be
x_m^s({\rm m.c.})=1\;.
\label{xmsmc}
\ee
The correlation length critical exponent $\nu$ is again obtained by
analysing the expression in eq(\ref{msdelta}). The typical number of return
points of the surviving walks is again $n=O(1)$, but now $l_i=O(L)$,
since the end-point of the walk is a return point. Consequently, the
correction term in eq(\ref{msdelta}) for surviving walks is $O(L)$, what
should be multiplied by the surviving probability to obtain the
average of $O(1)$, from which the micro-canonical correlation length
exponent
\be
\nu({\rm m.c.})=1\;,
\label{numc}
\ee
follows. Comparing the canonical and micro-canonical exponents in
eqs(\ref{xms}),(\ref{nu}) and eqs(\ref{xmsmc}),(\ref{numc}),
respectively, we can conclude that they are different. It is important,
however, to note that the surface magnetization exponent, defined via
\be
m_s(\delta,L\to\infty)\sim\delta^{\beta_s}
\ee
is the same for both ensemles: $\beta_s=x_m^s \nu=1$. In
Fig. \ref{fig5} we show the scaling plots for the surface
magnetization in the two ensembles obtained numerically by evaluating
eq(\ref{peschel}) for the binary distribution. Note that we expect
similar results to hold for end-to-end correlations
$[\langle\sigma_1^x\sigma_L^x\rangle]_{\rm av}$, with the exponent
$x_s$ replaced by $2x_s$ (c.f. sec. 3).

\begin{figure}
\epsfxsize=\columnwidth\epsfbox{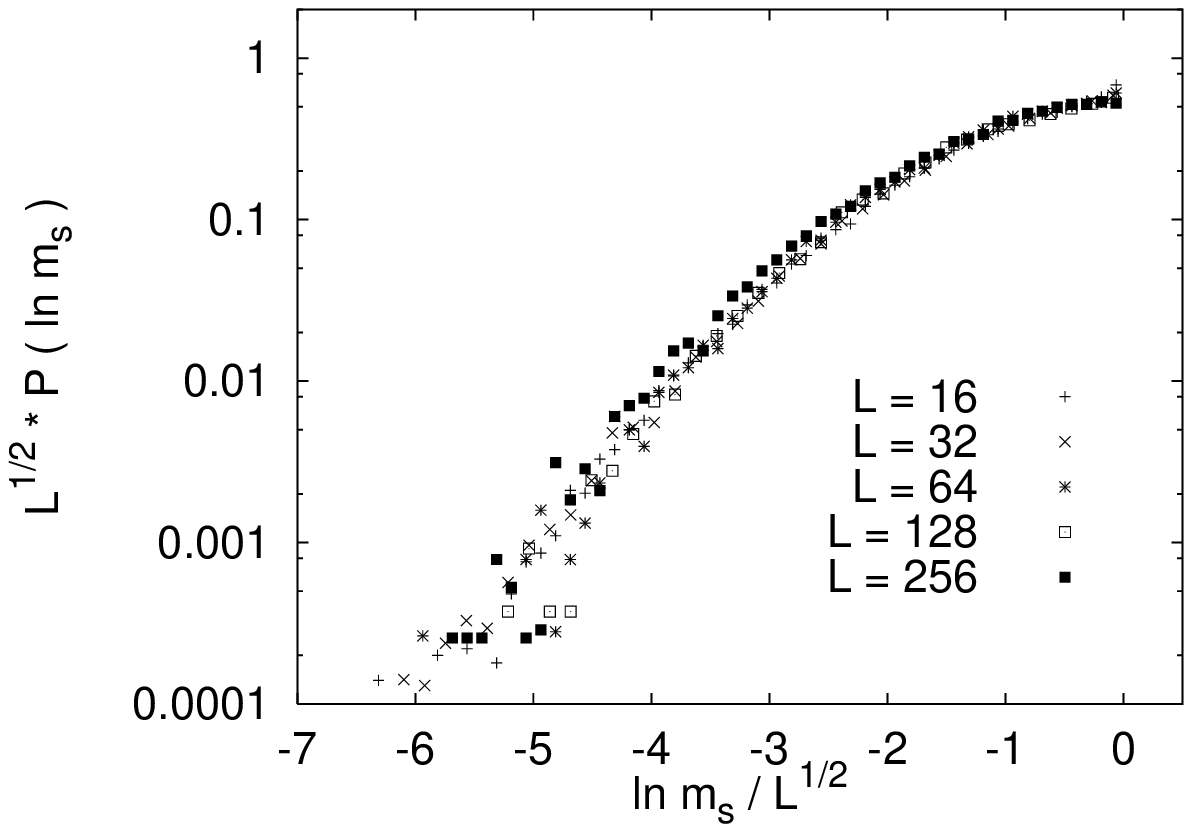}
\epsfxsize=\columnwidth\epsfbox{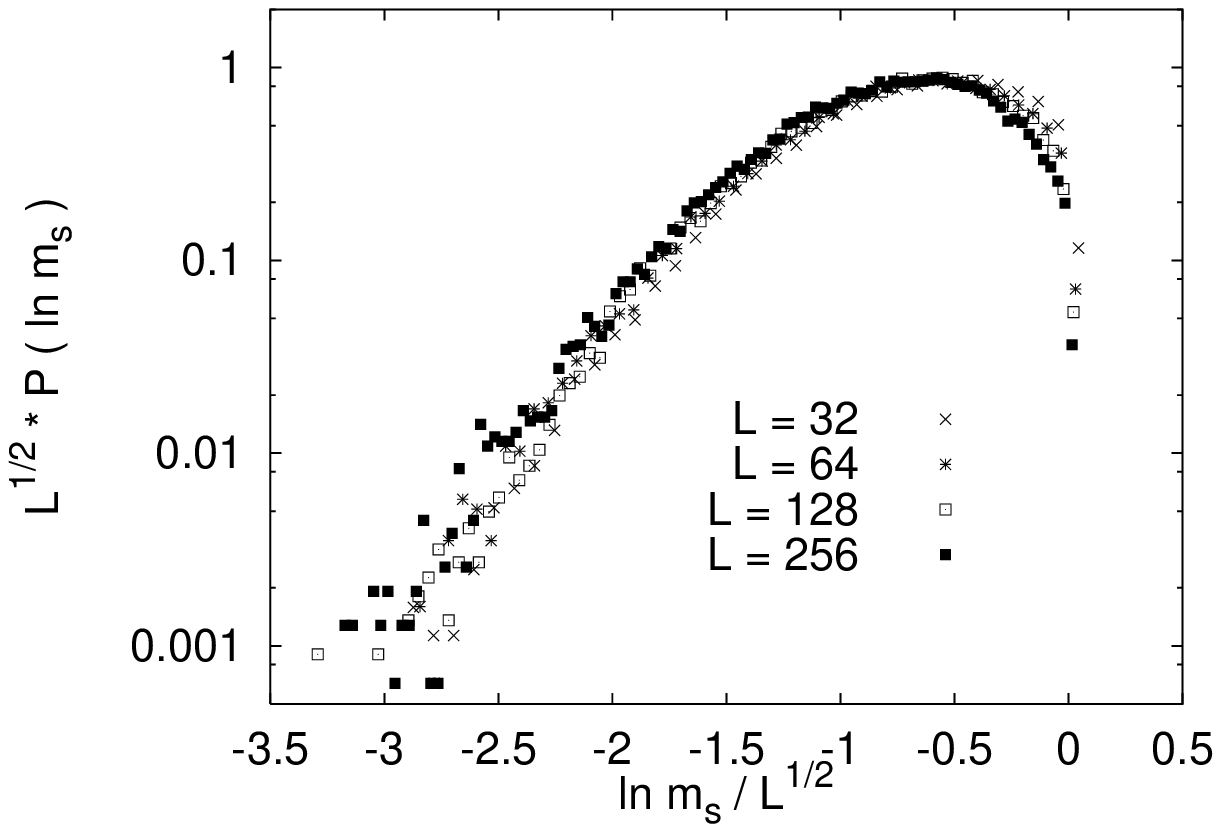}
\caption{Scaling plot of the probability distribution $P(\ln m_s)$ of
the surface magnetization in the canonical ensemble (top) and the
micro-canonical (bottom) ensemble. Both data are for the uniform distribution
averaged over 500000 samples.}
\label{fig6}
\end{figure}

The reason for the difference of the critical exponents measured in
the two ensembles is the fact that several physical quantities, among
those the surface magnetization, is {\it not self-averaging} at the
critical point. To illustrate this property in Fig. \ref{fig6} we have
plotted the probability distribution of the surface magnetization in
the two ensemles. Both scale, as expected, as
\be
P_L(\ln m_s) 
= \frac{1}{\sqrt{L}}\;\tilde{p}\left(\frac{\ln m_s}{\sqrt{L}}\right)
\ee
the asymptotic form of the scaling function $\tilde{p}$ for the
canonical ensemple was determined analytically \cite{fisher}, for
particular distributions of the fields and/or couplings it can even be
calculated exactly \cite{theo}. 

The average is determined by the rare events having a magnetization of
order $O(1)$, i.e.\ by the asymptotic behavior of $\tilde{p}(y)$ for
$y\to0$. From fig.\ \ref{fig5} (top) we conculde that for the
canonical ensemble $\tilde{p}(y)$ approaches a constant for $y\to0_-$,
whereas for the micro-canonical ensemble in fig.\ \ref{fig5} (bottom)
the scaling functions shows a power law dependence
\be
\tilde{p}(y)\sim (-y)^{a}
\qquad{\rm for}\quad y\to0_-
\ee
with a {\it positve} exponent $a$. For the average then follows
\beqn
[m_s]_{\rm av}&=&\frac{1}{\sqrt{L}}\int dm\;\tilde{p}(\ln m_s/\sqrt{L})\\
&\sim& \frac{1}{\sqrt{L}}\int dm\;(|\ln m_s|/\sqrt{L})^a 
\propto L^{-(1+a)/2}\nonumber
\eeqn
As we said above $a=0$ for the canonical ensemble, resulting in
$[m_s]_{\rm av}\sim1/\sqrt{L}$, and $a=1$ fits the data reasonably
well in case of the micro-canonical ensemble, resulting in $[m_s]_{\rm
av}\sim1/L$. 

Based on our observation on the surface magnetization we assume that
for other non self-averaging quantities the corresponding critical
exponents are generally ensemble dependent. In the rest of the paper
we restrict ourselves to the canonical ensemble.

\subsection{Profiles of observables}

A real system is always geometrically constrained and due to modified
surface couplings its properties in the surface region are generally
different from those in the bulk. Close to the critical point this
surface region, which has a characteristic size of the correlation
length, intrudes far into the system. At the very critical point the
appropriate way to describe the position dependent physical quantities
is to use density profiles rather then bulk and surface
observables. For a number of universality classes much is known about
the spatially inhomogeneous behavior, in particular in two-dimensions,
where conformal invariance provides a powerful tool to study various
geometries \cite{ipt}.

In a critical system confined between two parallel plates, which are at a
large, but finite distance $L$ apart, the local densities $\Phi(r)$
such as the order parameter (magnetization) or energy density vary
with the distance $l$ from one of the plates as a smooth function of
$l/L$. According to the scaling theory by Fisher and de
Gennes \cite{fisherdegennes}:
\be
\langle\Phi(l)\rangle_{ab}=L^{-x_{\Phi}}F_{ab}(l/L)\;,
\label{fisherdegennes}
\ee
where $x_{\Phi}$ is the bulk scaling dimension of the operator $\Phi$,
while $ab$ denotes the boundary conditions at the two plates. In a
$d$-dimensional system the scaling function in eq(\ref{fisherdegennes})
has the asymptotic behavior:
\be
F_{ab}(l/L)=A\left[ 1 + B_{ab} \left({l \over L}\right)^d+\dots \right]
~~~{l \over L}\ll 1\;.
\label{fab}
\ee
Here the amplitude of the first correction term is universal, the
corresponding exponent is just $x_{\Phi}^s=d$ the surface scaling
dimension of $\Phi$.

In two-dimensions conformal invariance gives further predictions on the
profile:
\be
\langle\Phi(l)\rangle_{ab}=\left[{L\over \pi} \sin \pi{l \over L} \right]^{-x_{\Phi}}
G_{ab}(l/L)\;,
\label{confprof}
\ee
where the scaling function $G_{ab}(l/L)$ depends on the universality class of
the model and on the type of the boundary condition. With symmetric boundary
conditions the scaling function is constant $G_{aa}=A$. For conformally
invariant, non-symmetric boundary conditions the scaling function has been
predicted for several models. For the Ising model the magnetization profiles
with free-fixed (f+) and (+-) boundary conditions are predicted as:
\be
G_{f+}=A\left[\sin {\pi l \over 2 L}\right]^{x_m^s}\;,
\label{isingf+}
\ee
and
\be
G_{+-}=A \cos{\pi l \over L}\;,
\label{ising+-}
\ee
respectively.

\begin{figure}
\epsfxsize=\columnwidth\epsfbox{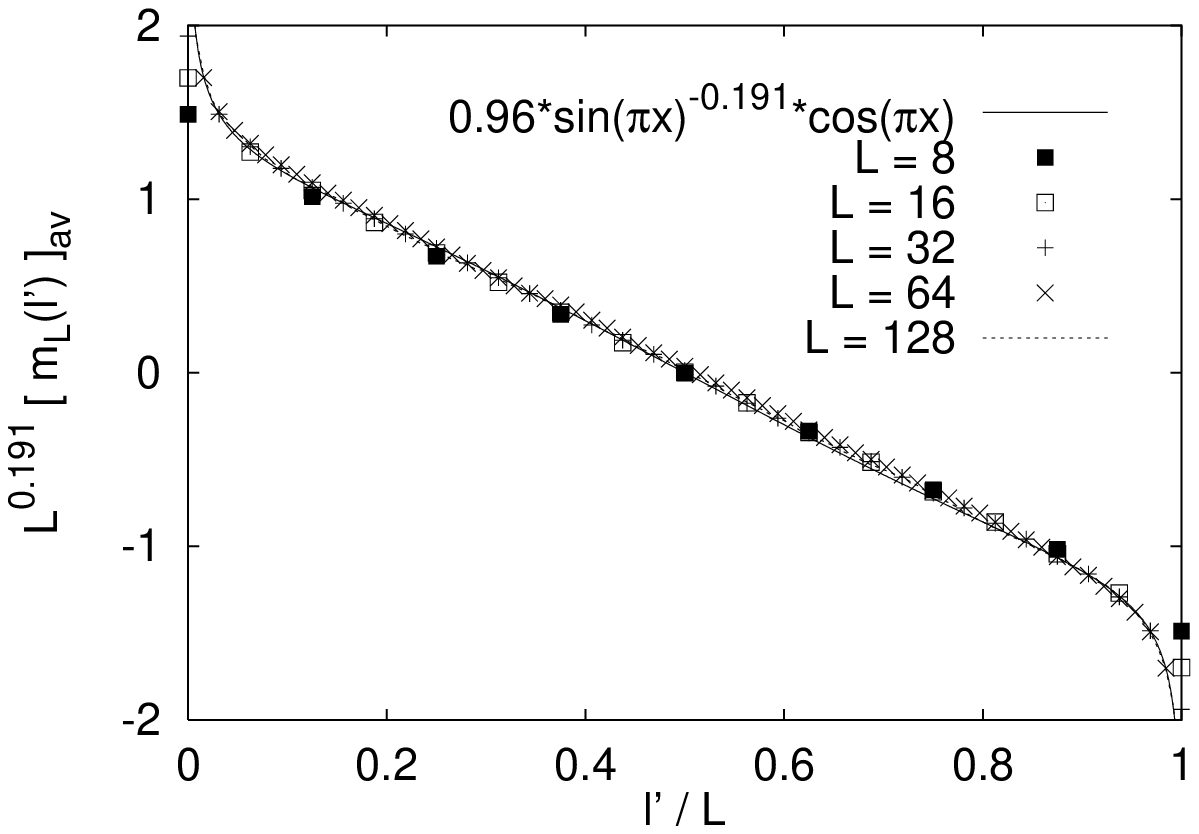}
\epsfxsize=\columnwidth\epsfbox{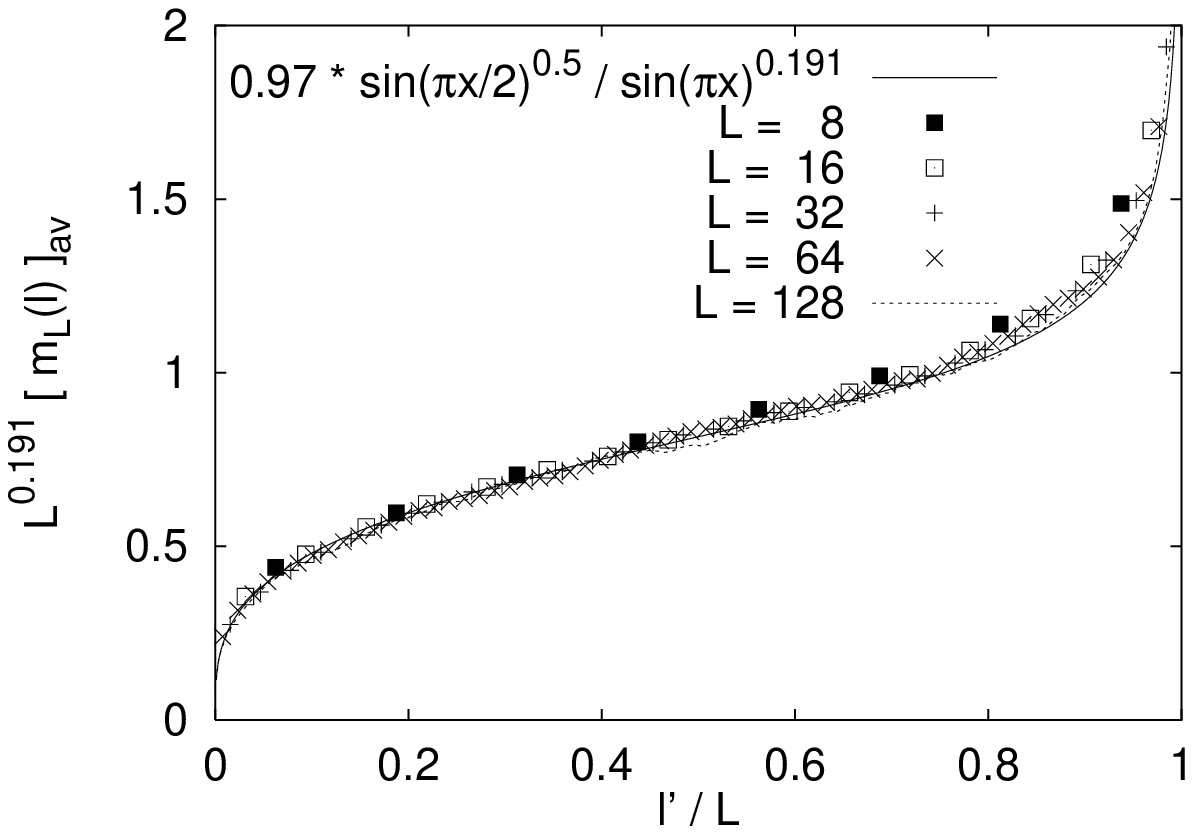}
\caption{Scaling plots of the magnetization profiles for non-symmetric
boundary conditions ($l'=l-0.5$). Top: plus-minus ($+-$) b.c, the
broken line is a fit to the form (\protect{\ref{confprof}}) and
(\protect{\ref{ising+-}}) with $A$ as a fit parameter. Bottom:
free-fixed (f+) b.c., the broken line is a fit to the form
(\protect{\ref{confprof}}) and (\protect{\ref{isingf+}}) with $A$ as a
fit parameter.}
\label{fig7}
\end{figure}

In two-dimensions conformal invariance can also be used to predict the critical
off-diagonal matrix-element profiles $\langle\Phi|\Phi(l)|0\rangle$, where $\langle\Phi|$
denotes the lowest excited state leading to a non-vanishing matrix-element
(see eq(\ref{magnfree})). These off-diagonal profiles give information about
the surface and bulk critical behavior via finite-size scaling while avoiding
the contribution of regular terms. With symmetric boundary conditions one
obtains for the profile \cite{turbanigloi97}:
\be
\langle\Phi|\Phi(l)|0\rangle \propto \left({\pi \over L}\right)^{x_{\phi}}
\left(\sin \pi {l \over L} \right)^{x_{\Phi}^s-x_{\Phi}}\;,
\label{offdiag}
\ee
which involves both the bulk and surface scaling dimensions.

\begin{figure}
\epsfxsize=\columnwidth\epsfbox{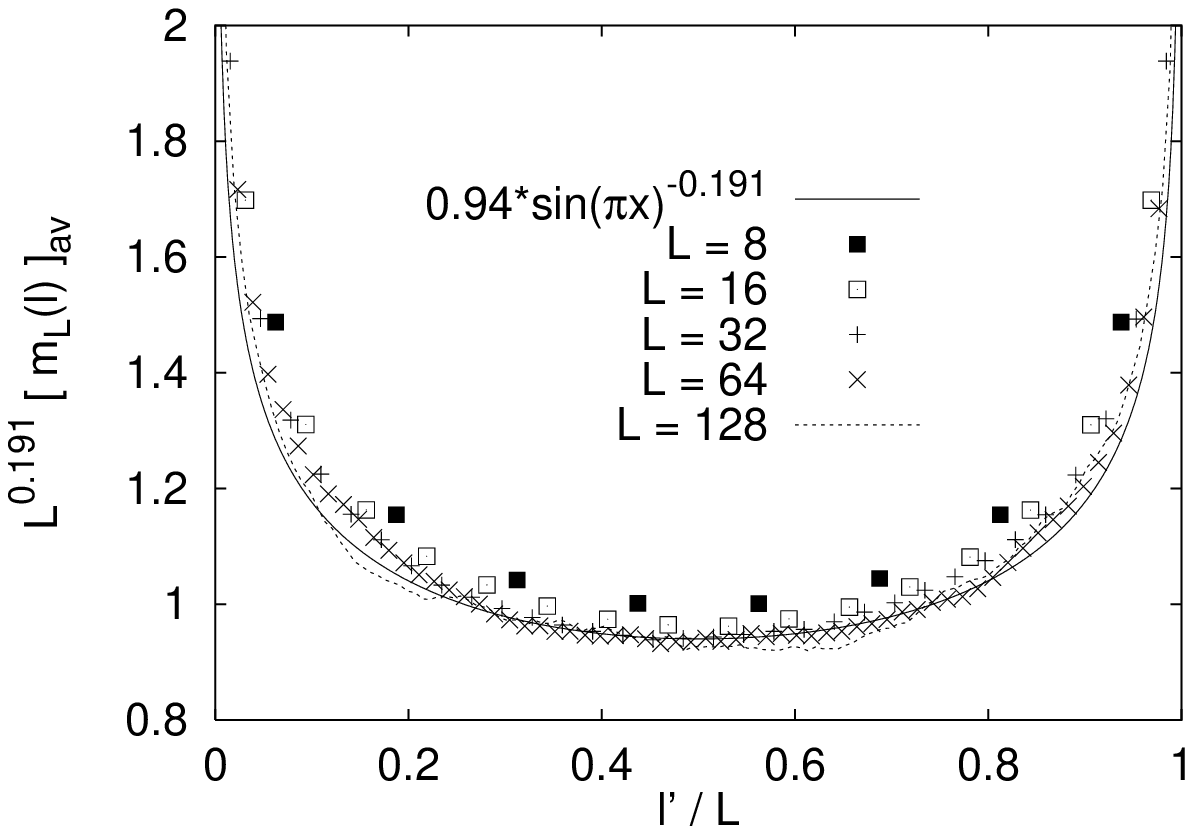}
\epsfxsize=\columnwidth\epsfbox{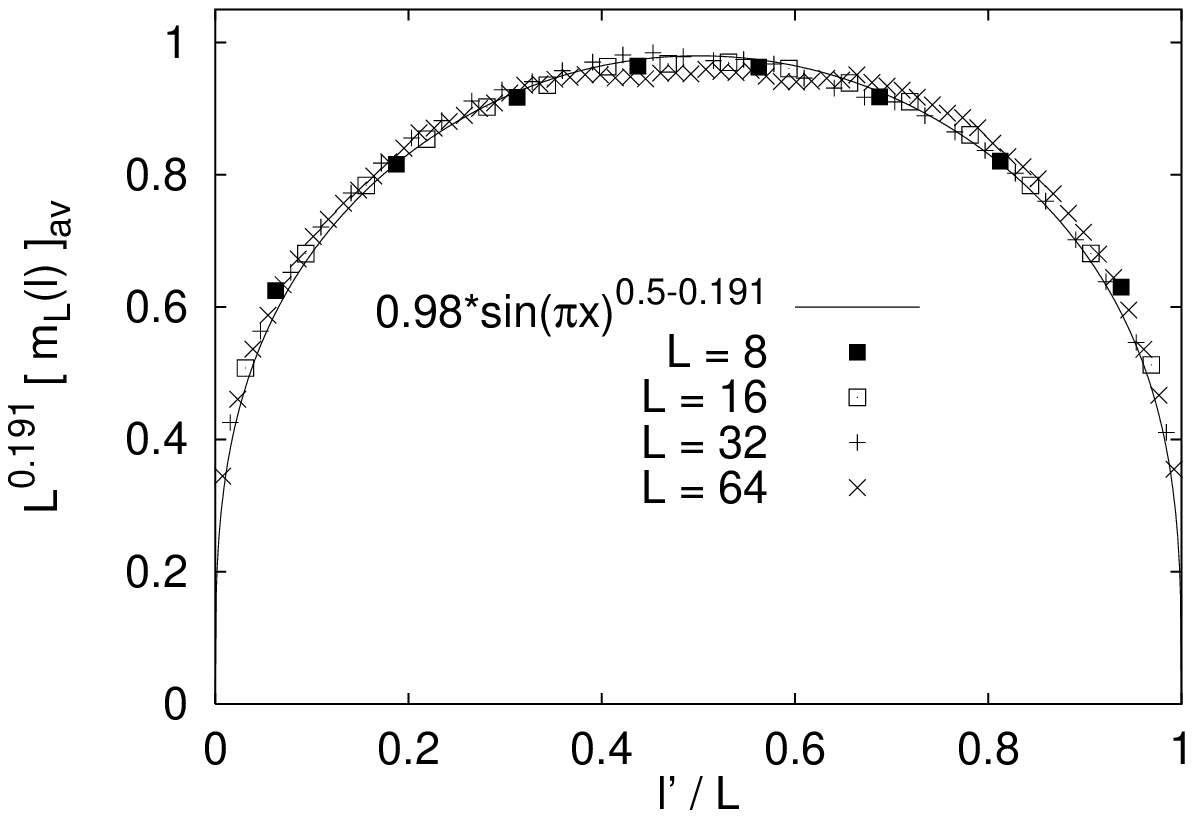}
\caption{Top: scaling plot of the magnetization profile for
symmetric (here: fixed) boundary conditions. The broken line is a fit
to the form (\protect{\ref{confprof}}) and $G_{aa}=A$ with $A$ as a
fit parameter. Bottom: the profile of the off-diagonal matrix element
with free b.c. The broken line is a fit to the form
(\protect{\ref{offdiag}}).}
\label{fig8}
\end{figure}

The numerically calculated average diagonal and off-diagonal
magnetization profiles (see sec.\ 2) for the RTIM are presented in
Fig. \ref{fig7} and \ref{fig8} for the uniform distribution (in
\cite{profiles} some data for the binary distribution have been
presented). Here we do not use $x_m$ and $x_m^s$ as fit parameters,
but fix them to the theoretically predicted values in eqs(\ref{beta}),
(\ref{xms}). The only fit parameter is the non-universal prefactor
$A$, which is found remarkably constant for the different boundary
conditions.  As one can see on Fig. \ref{fig7} the data for different
length $L$ collapse to scaling curves, which are very well described
by the scaling functions predicted by conformal invariance. Thus we
can conclude that not only the scaling prediction by Fisher and de
Gennes\cite{fisherdegennes} in eq(\ref{fisherdegennes}) is very well
satisfied for the RTIM, but the corrections to the appropriate
conformal results are also very small, practically negligible. This is
an unexpected result, since the RTIM is not conformally invariant, due
to anisotropic scaling at the critical point eq(\ref{scales}).

To close this section we present numerical results for the bulk
magnetization scaling dimension $x_m$ and compare it with Fisher's
perhaps most striking prediction in eq(\ref{beta}). Here we have made
effort to increase the numerical accuracy, therefore we worked with
the binary distribution in eq(\ref{binary}) on chains with both ends
fixed with length $L \le 24$ and performed the {\it exact average} of
the local magnetization on the central spin. From the finite-lattice
magnetizations, which scales as $[m(L,0)]_{av} \sim L^{-x_m}$ we have
determined $x_m$ by two-point fit, comparing systems with sizes $L$
and $L-2$. From the finite-size exponents, presented in Table 1 for
$\lambda=2,3$ and 4 one conclude, that they are in agreement with
Fisher's result: $x_m=(3-\sqrt{5})/4=.191$. Unfortunatelly the
numerical data in Table 1 show log-periodic oscillations, which is a
consequence of the energy scale introduced by the binary
distribution. Therefore one can not use the accurate
sequence-extrapolation methods to analyse the limiting behavior of the
series.  Then from a simple linear fit one can obtain the estimate:
\be
x_m=0.190 \pm 0.003\;,
\label{xmest}
\ee
improving the accuracy of previous MC estimates\cite{youngrieger}.
%
\bc
\begin{tabular}{|c||c|c|c|}
\hline
$\quad$L$\quad$ & \multicolumn{3}{c|}{$x_m(L)$}\\
\hline
& $\quad\lambda=2\quad$ & $\quad\lambda=3\quad$ & $\quad\lambda=4\quad$\\
\hline
6&.127071&.162136&.181770\\
8&.142310&.161044&.169656\\
10&.157063&.179177&.189815\\
12&.167197&.195090&.207268\\
14&.173605&.197072&.206820\\
16&.176458&.196602&.204265\\
18&.178444&.195288&.201673\\
20&.179836&.194391&.199992\\
22&.181044&.194279&.199270\\
24&.182175&&\\
\hline
\end{tabular}
\ec
TABLE I: Numerical estimates for the bulk magnetization exponent
$x_m(L)$ for the binary distribution for various values of $\lambda$.
%
%

\subsection{Dynamical correlations}

The general time and position dependent correlations in eq(\ref{gencorr})
have a complicated structure at the critical point. Therefore we consider
dynamical correlations on the same spin, which has a simpler asymptotic
behavior. First we consider the bulk autocorrelation function 
\be
G(\tau)=[\langle\sigma_{L/2}^x(\tau)\sigma_{L/2}^x\rangle]_{\rm av}
\label{corrdef}
\ee
and recapitulate the scaling argument in Ref\onlinecite{riegerigloi}.

The autocorrelation function, like to the (local) magnetization, is not
self-averaging at the critical point: its average value is determined
by the {\it rare events}, which occour with a probability $P_r$ and
$P_r$ vanishes in the thermodynamic limit. In the random quantum systems
the disorder is strictly correlated along the time axis, consequently
in the rare events with a local order, i.e. with a finite magnetization also
the autocorrelations are non-vanishing. Under a scaling transformation, when
lengths are rescaled as $l'=l/b$, with $b>1$ the probability of the rare
events transforms as $P_r'=b^{-x_m}$, like to the local magnetization. As we
said above the same is true for the autocorrelation function:
\be
G(\ln \tau)=b^{-x_m} G(\ln \tau /b^{1/2})~~~\delta=0\;,
\label{crscale}
\ee
where we have made use of
the relation between relevant time $t_r$ and length $\xi$ at the
critical point in eq(\ref{scales}). Taking now the length scale as
$b=(\ln \tau)^2$ we obtain:
\be
G(\tau) \sim (\ln \tau)^{-2x_m}~~~\delta=0\;.
\label{crcorr}
\ee
For surface spins in eqs(\ref{crscale},\ref{crcorr}) the surface
magnetization scaling dimension $x_m^s$ appears.

\begin{figure}
\epsfxsize=\columnwidth\epsfbox{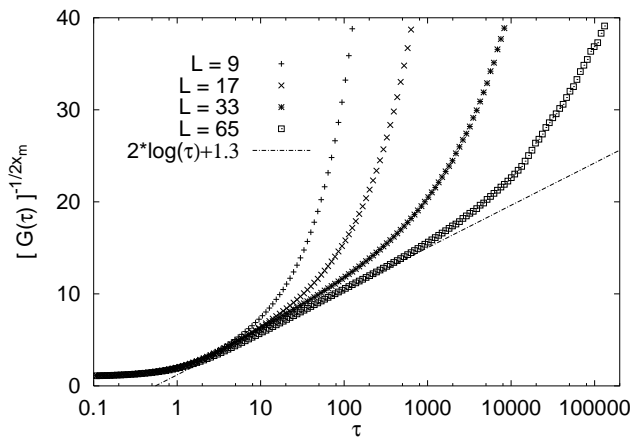}
\caption{Bulk spin-spin autocorrelation function 
  $G(\tau)$ eq(\protect{\ref{corrdef}})
  for various system sizes (and the uniform
  distribution). The straight line is the prediction according to
  eq(\protect{\ref{crcorr}})}.
\label{fig9}
\end{figure}

In Fig. \ref{fig9} we present the numerical results for the critical
bulk autocorrelation function obtained via evaluating the Pfaffian
eq(\ref{pfaffian}). Note that we have chosen $L$ to be odd, so that
$L/2$ denotes the central spin, representing the bulk behavior in a
system with free b.c. A plot of $G(\tau)^{-1/2x_m}$, with $x_m$ as in
(\ref{xmest}), versus $\ln \tau$ (or $\tau$ on a logarithmic scale)
should yield a straight line in the infinite system size limit
according to eq(\ref{crcorr}). As can be seen in Fig.  \ref{fig9} the
data agree well with this prediction. For the surface autocorrelations
$G_1(\tau)=[\langle\sigma_1^x(\tau)\sigma_1^x\rangle]_{\rm av}$,
evaluated according to eq(\ref{surfcorr}), which is much less involved
than the computation of a pfaffian, a similar plot with the bulk
magnetization exponent $x_m$ replaced by the surface magnetization
exponent $x_m^s$ gives also an excellent agreement with the prediction
$G_1(\ln\tau)\sim(\ln\tau)^{-2x_m^s}$.

To complete the results on critical dynamics we mention the scaling behavior
of the energy-energy autocorrelation function $[\langle \sigma_l^z(\tau)
\sigma_l^z\rangle]_{av}$. As shown in Ref\onlinecite{riegerigloi} this
quantity is self-averaging at the critical point and can be characterised
by a power law asymptotic decay with novel critical exponents, which are
different in the bulk and at the surface of the system.

\section{Off-critical properties}

A surprising property of random quantum systems is the existence of
Griffiths-McCoy singularities in the paramagnetic side of the critical
point. In the corresponding Griffiths-McCoy region the autocorrelation
function decays as a power $G(\tau) \sim \tau^{-1/z(\delta)}$, where the
dynamical exponent $z(\delta)$ characterises also the distribution of
low energy excitations in eq(\ref{gapdistr}). As a consequence, the
free energy is non-analytic function of the magnetic field and the
susceptibility diverges in the whole region.

According to the phenomenological theory\cite{youngrieger} in the
Griffiths-McCoy region the singularities of all physical quantities
are entirely characterised by the dynamical exponent $z(\delta)$.
Numerical calculations\cite{youngrieger,young} give support to this
assumption, although there are discrepancies between the values of
$z(\delta)$ obtained from different quantities.

Here we extend previous investigations in several respects. First, we
consider also the surface properties, such as the surface
autocorrelation function and the surface susceptibility. Second, we
investigate also the {\it ferromagnetic side} of the critical point.
In the neighborhood of the critical point Fisher\cite{fisher} has
already obtained some RG results in the ferromagnetic phase. Here we
are going to check these results numerically and to extend them for
finite $\delta<0$.

\subsection{Phenomenological scaling considerations}

As already shown in Section 4 the dynamical exponent $z(\delta)$ is
conveniently measured from the probability distribution of the energy
gap (in the ferromagnetic phase one considers the second gap in a
finite system, which does not vanish exponentially.) For large systems
the gap-distribution is given by eq(\ref{gapdistr}) and with this the
average autocorrelation function is given by:
\be
G(\tau) \sim \int_0^{\infty} P(\epsilon) \exp(-\tau \epsilon) d\epsilon
\sim \tau^{-1/z(\delta)}\;.
\label{autocorr1}
\ee
In a finite system of size $L$ for long enough time, such that
$\tau \gg L^{z(\delta)}$, the decay in eq(\ref{autocorr1}) will change
to a $G(\tau) \sim 1/\tau$ form, which is characteristic for isolated
spins. It means that in the above limit the system can be considered as
an effective single spin.

In the following we present a simple scaling theory which explains the
form of the asymptotic decay in eq(\ref{autocorr1}).
Here in the Griffiths-McCoy phase we modify the scaling relation in
eq(\ref{crscale}) by two respects. First, the scaling combination is changed to
$\tau/b^z$, since the dynamical exponent $z(\delta)$ is finite in the
off-critical region. Second, the {\it rare events}, which are
responsible for the Griffiths-McCoy singularities are now samples with
very low energy gaps and their number is practically independent of
the size of the system. Consequently the rescaling prefactor is
$b^{-1}$ and the scaling relation is given by:
\be
G(\tau,1/L)=b^{-1} G(\tau/b^z,b/L)~~~\delta\ne 0\;,
\label{offscale}
\ee
where the inverse size of the system $1/L$ is also included as a scaling
field. Now taking $b=\tau^{1/z}$ we obtain:
\be
G(\tau,1/L)=\tau^{-1/z} \tilde{G}(\tau^{1/z}/L)~~~\delta \ne 0\;, 
\label{offcorr}
\ee
thus in the thermodynamic limit we recover the power law decay in
eq(\ref{autocorr1}). The scaling function $\tilde{G}(y)$ in
eq(\ref{offcorr}) should behave as $\tilde{G}(y) \sim y^{1-z}$ for
large $y$, in this way one recovers the limiting $1/\tau$ decay, as
argued below eq(\ref{autocorr1}). Then the
finite size scaling behavior of the autocorrelation function is of the
form of $L^{z-1}$, and after integrating $G(\tau,1/L)$ by $\tau$ the same
scaling behavior will appear in the local susceptibility:
\be
\chi_i(L) \sim L^{z-1}~~~\delta \ne 0\;.
\label{locsusc2}
\ee

\begin{figure}
\epsfxsize=\columnwidth\epsfbox{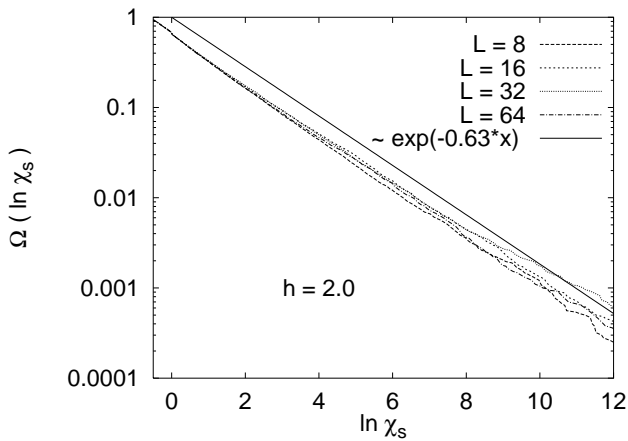}
\caption{Integrated probability distribution of the zero frequency
surface susceptibility for different system sizes in the disordered
phase at $h=2.0$. Note that $z(h)$, as determined from the slope of
the straight line, turns out to be within the error margin of $z(h)$
determined via the gap distribution (see Fig. \protect{\ref{fig2}}).}
\label{fig10}
\end{figure}

\subsection{Numerical calculation of the dynamical exponent}

The phenomenological description of the Griffiths phase suggests that
all Griffiths-McCoy singularities emerging in temperature, energy,
time or frequency dependent quantities should be paramtrizable by a
single dynamical expontent $z(\delta)$. In this subsection we present
the results on our numerical estimates for $z(\delta)$ resulting from
the calculation of the following quantities:
\begin{itemize}
\itemsep0cm
\parsep0cm
\item distribution of low energy excitations,
\item autocorrelation function on bulk and surface spins,
\item distribution of surface susceptibilities.
\end{itemize}
\begin{figure}
\epsfxsize=\columnwidth\epsfbox{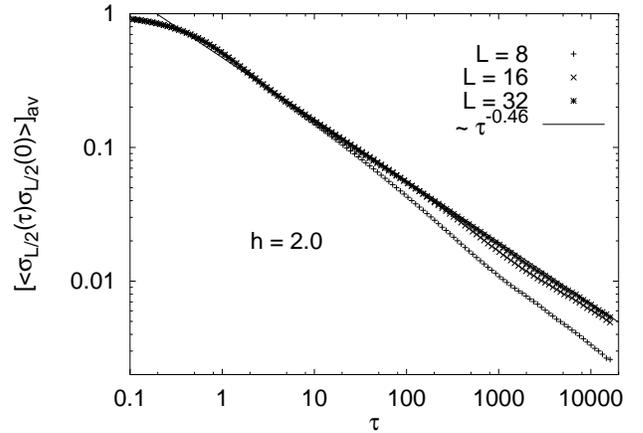}
\epsfxsize=\columnwidth\epsfbox{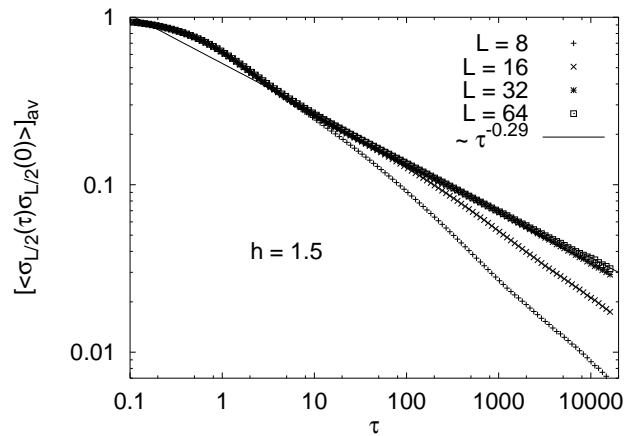}
\caption{The bulk autocorrelation function
$[\langle\sigma^x_{L/2}(\tau)\sigma^x_{L/2}(0)\rangle]_{\rm av}$ in
imaginary time in the disordered phase ($h>1$), calculated at a
central spin $i=L/2$ with eq(\protect{\ref{gencorr}}) via the Pfaffian
method presented in section II. The straight lines are fits to the
expected power law decay $\tau^{-1/z(h)}$.}
\label{fig11}
\end{figure}
The distribution functions for the energy gaps have already been
presented in Section 4. The same quantity for the surface
susceptibility in eq(\ref{surfsusc}) has a similar form as the inverse
gap, as seen in Fig. \ref{fig10}. The only difference that for the
susceptibility the matrix-element in the denominator of
eq(\ref{surfsusc}) select one special position of the SCD. As a
consequence the corresponding probability distribution has no
$L$-dependence, as already discussed in\cite{youngrieger} and can also be
seen by comparing eq(\ref{epspara}) with (\ref{locsusc2}). The
$z(\delta)$ exponents calculated from the surface susceptibility
distribution agree well with those obtained from the gap distribution.

\begin{figure}
\epsfxsize=\columnwidth\epsfbox{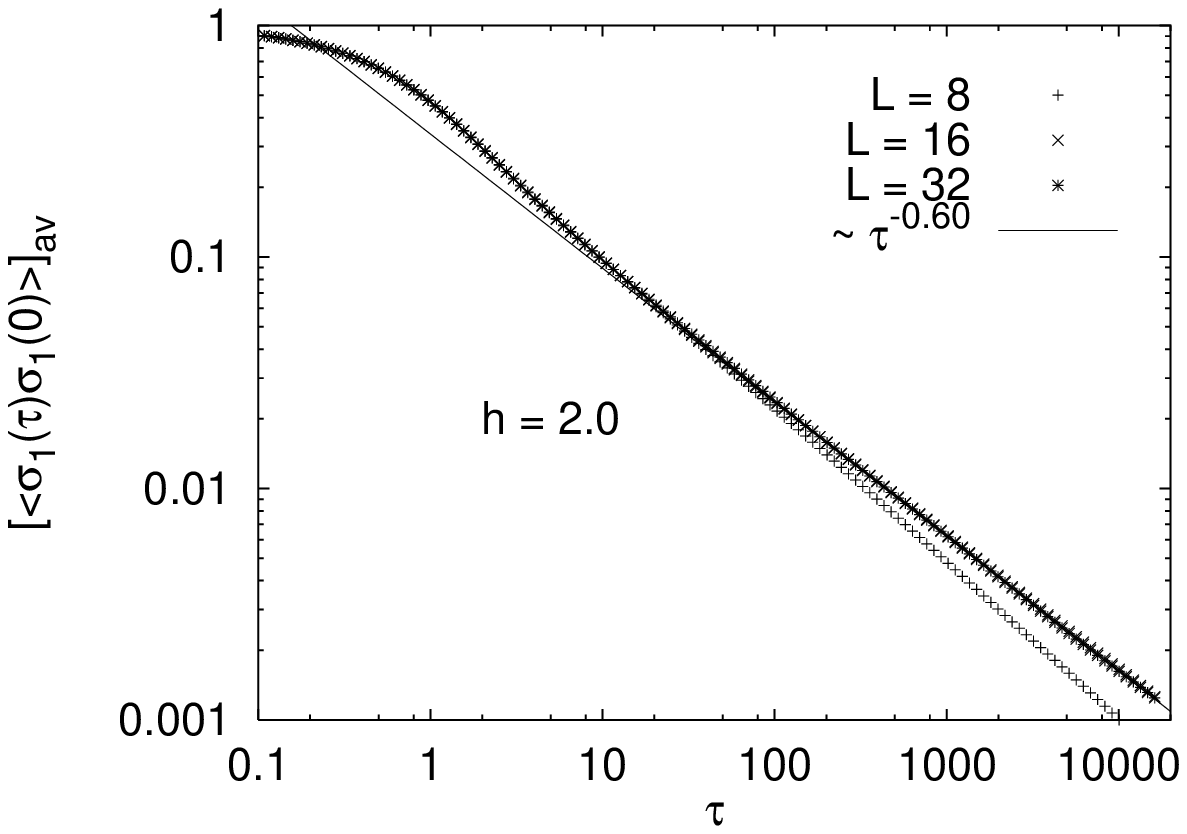}
\epsfxsize=\columnwidth\epsfbox{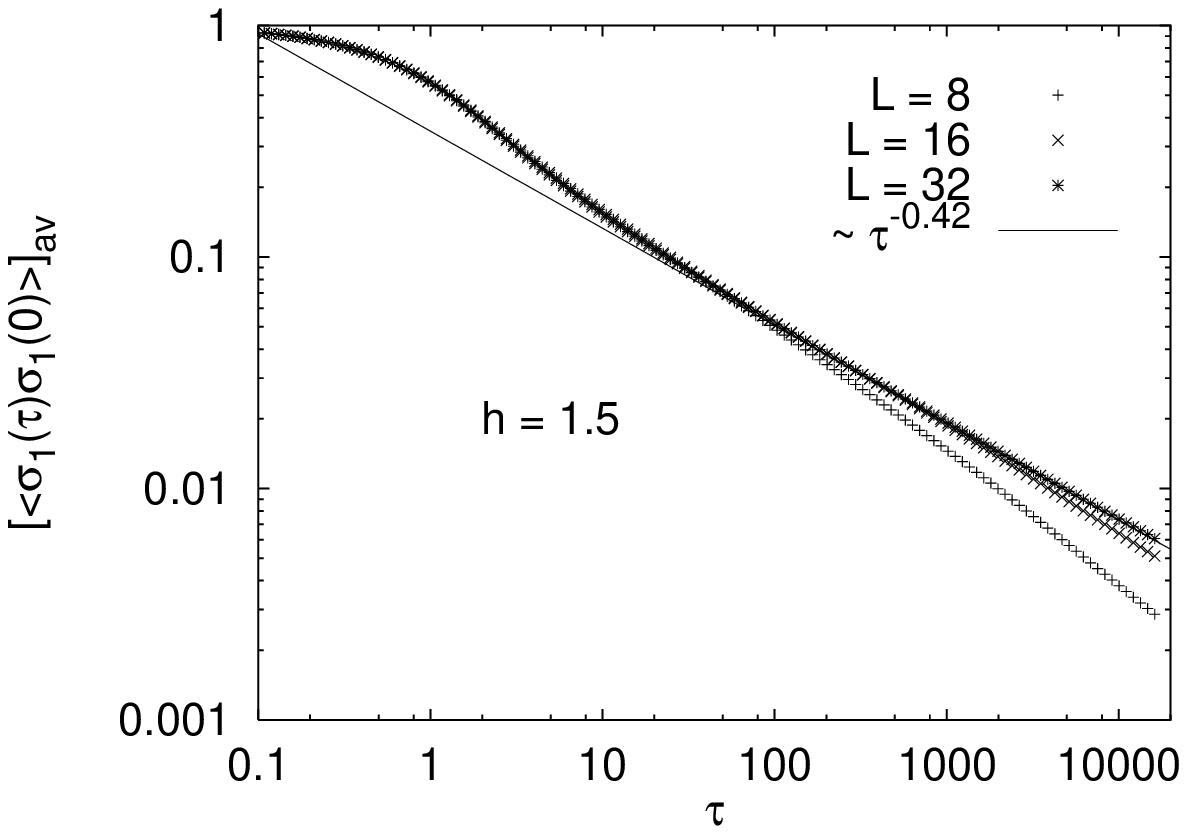}
\caption{The surface autocorrelation function
$[\langle\sigma^x_{1}(\tau)\sigma^x_{1}(0)\rangle]_{\rm av}$ in imaginary
time in the disordered phase ($h>1$), calculated via
eq(\protect{\ref{surfcorr}}). The straight lines are fits to the
expected power law decay $\tau^{-1/z(h)}$.}
\label{fig12}
\end{figure}

\begin{figure}
\epsfxsize=\columnwidth\epsfbox{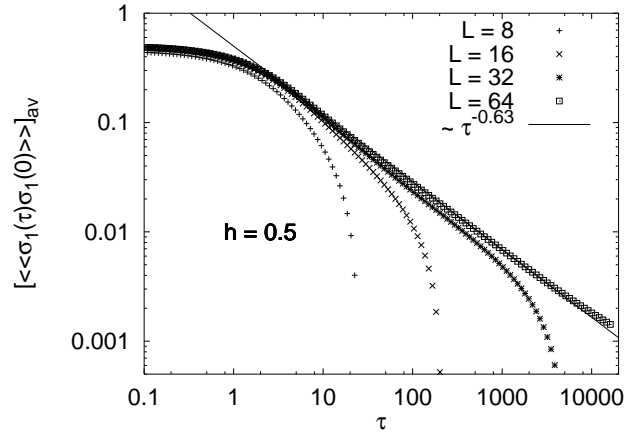}
\caption{The connected part of the surface autocorrelation function
$[\langle\langle\sigma^x_{1}(\tau)\sigma^x_{1}(0)\rangle\rangle]_{\rm
av}=[\langle\sigma^x_{1}(\tau)\sigma^x_{1}(0)\rangle-m_s^2]_{\rm av}$ in
imaginary time in the ordered phase ($h<1$), calculated via
eq(\protect{\ref{surfcorr}}), but now substracting $|\Phi_1(1)|^2$. The
straight lines are fits to the expected power law decay
$\tau^{-1/z(h)}$.}
\label{fig13}
\end{figure}

The average autocorrelation function is measured at two sites of the
chain: on the central spin, giving an estimate for the bulk
correlation function and on the surface spin. The average bulk
autocorrelation functions are drawn on a log-log plot in
Fig. \ref{fig11} for several values of $\delta>0$.  One can easily
notice an extended region of the curves, which are well approximated
by straigth lines, the slope of which is connected to the dynamical
exponent through eq(\ref{autocorr1}).  Similar behavior can be seen on
Fig. \ref{fig12}, where the average surface autocorrelation functions
are drawn.  Our investigation on the dynamical exponent is completed
by studying the {\it connected} surface autocorrelation function in
the ferromagnetic phase. As seen on Fig. \ref{fig13} the scaling form
in eq(\ref{autocorr1}) is well satisfied for this function, too.

\begin{figure}
\epsfxsize=\columnwidth\epsfbox{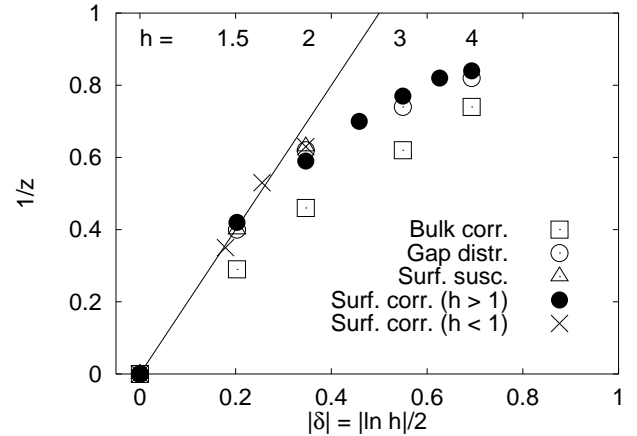}
\caption{Summary of all estimates of the dynamical exponent $z(h)$ as
a function of the distance from the critical point $\delta=|\ln h|/2$.
The open circles are the estimates from the gap distribution, as
exemplified in Fig. \protect{\ref{fig2}}, open tiangles: surface
susceptibility distribution (c.f.\ Fig. \protect{\ref{fig10}}), full
circles: surface autocorrelation function (c.f.\
Fig. \protect{\ref{fig12}}), crosses: surface autocorrelation function
in the ordered phase (c.f.\ Fig. \protect{\ref{fig13}}), open squares:
bulk autocorrelation function (c.f.\ Fig. \protect{\ref{fig11}}). Note
that wheras all former estimates agree within the error margin (which
is roughly the size of the symbols) the latter estimate, namely the
one obtained via the bulk autocorrelation function, differs
significantly from all others.}
\label{fig14}
\end{figure}

The behavior of the dynamical exponents calculated by different
methods are summarized in Fig. \ref{fig14}. First we note that the
numerical estimates are very close to each other. The only exception
is the data obtained from the bulk autocorrelations. To explain the
possible origin of this discrepancy we turn in the Discussion. The
$z(\delta)$ values well satisfy the two theoretical limits:
$\lim_{|\delta| \to \infty} z(\delta)=1$ and $\lim_{|\delta| \to
  0}=1/2\delta$\cite{fisher}. Furthermore the dynamical exponents show
the duality relation: $z(\delta)=z(-\delta)$.

\section{Discussion}

In this paper the critical and off-critical properties of the random
transverse-field Ising spin chain are studied by analytical and
numerical methods and by phenomenological scaling theory. The
previously known exact\cite{mccoywu}, RG\cite{fisher} and numerical
results\cite{youngrieger,profiles,riegerigloi,young} about the model
have been extended and completed here in several directions. The
scaling behavior of the surface magnetization is obtained through a
mapping to an adsorbing random walk and the critical exponents
$\beta_s$, $\nu$ and $x_m^s$ are calculated exactly. We have also
shown that the values of these exponents depend on the ensemble over
which the averaging is performed.

Using the correspondence between the surface magnetization and the
adsorbing walks we have identified strongly coupled domains in the
system, where the couplings have a surviving walk character, and
estimated the distribution of low energy excitations both in the
critical and off-critical regions. This provides a comprehensive
explanation of the microscopic origin of the Griffiths-McCoy
singularities. It turns out that most of the astonishing features of
the critical as well as the off-critical (Griffiths-McCoy) properties
can be simply explained via random walk analogies. However, one
prediction by Fisher \cite{fisher}, namely the exact value of the bulk
magnetization exponent $\beta$ and its surprising relation to the
golden mean, still lacks a {\it simple} explanation in terms of
universal properties of random walks.

In the numerical part of our work we have treated relatively large $(L
\le 128)$ finite systems. At the critical point we have calculated the
magnetization profiles for different boundary conditions, which are
found to follow accurately the conformal predictions, although the
system is not conformally invariant. We have also increased the
numerical accuracy in the calculation of the bulk magnetization
scaling dimension. In the off-critical regions we have determined the
dynamical exponent $z(\delta)$ from different physical quantities. The
obtained results give support to the scaling prediction that the
Griffiths-McCoy singularities are characterised by the single
parameter $z(\delta)$. Here we note that the numerical data show
systematic differencies, when $z(\delta)$ is calculated from bulk or
from surface quantities. Similar observation has been made in
ref\cite{young}, too. The possible origin of the discrepancies is,
that the SCD-s, which are responsible for the Griffiths-McCoy
singularities, have different environments at the surface and in the
volume of the system.  Then, from the argument leading to
eqs(\ref{lpara},\ref{epspara}) one can obtain logarithmic corrections
between the dynamical exponents. This fact can then explain the
differences in the finite-size data. We have here, at the first time,
numerically studied the Griffiths-McCoy singularities in the
ferromagnetic phase, too. In this region the second gap of the
Hamiltonian and the {\it connected} autocorrelation function scale
with the dynamical exponent, which, according to numerical results,
satisfies the duality relation.

\acknowledgements 
F.\ I.'s work has been supported by the Hungarian National Research
Fund under grants No OTKA TO12830, OTKA TO23642, OTKA TO17485 and OTKA
TO15786 and by the Ministery of Education under grant No FKFP
0765/1997. We are indebted to L. Turban for a critical reading of the
manuscript. Useful discussions with T. Nieuwenhuizen, F. Pazmandi, G.
Zimanyi and A. P. Young are gratefully acknowledged.  H.\ R.'s work
was supported by the Deutsche Forschungsgemeinschaft (DFG) and he
thanks the Aspen Center for Physics, the International Center of
Theoretical Physics in Trieste and the Research Institute for Solid
State Physics, Budapest, where part of this work has been completed,
for kind hospitality.

\appendix
\section*{Adsorbing random walks}

Here we summarize the basic properties of one-dimensional random walks
in the presence of an adsorbing wall. For simplicity, first we consider a
walker, which makes steps of unit lengths with probabilities $p$ and
$q=1-p$ to the positive and to the negative directions, respectively.
Starting at a distance $s>0$ from an adsorbing wall we are interested in
the surviving probability $P_{\rm surv}(\delta_w,L)$ after $L$ steps. Here
\be
\delta_w=q-p\;
\label{deltaw}
\ee
measures the average drift of the walk in one step: for $\delta_w<0$
($\delta>0$) the walk has a drift towards (off) the wall.

The probability $W_L(l)$, that the walker after $L$ steps is at a position
$l \le L$, can be easily obtained by the mirror method\cite{physstat}:
\beqn
&&W_L(l)=\label{WLl}\\
&&p^{L+l \over 2} q^{L-l \over 2} \left({L! \over \left({L+l \over 2}
\right)!\left({L-l \over 2} \right)!} -{L! \over \left({L+l \over 2}
+s\right)!\left({L-l \over 2} -s \right)!} \right)\;.\nonumber
\eeqn
In the following we take $s=1$ and in the limit $L \gg 1$, $l \gg 1$ we
use the central limit theorem to write in the continuum approximation
$l \to x$ and $W_L(l) \to P_L(x)$ as:
\be
P_L(x)={1 \over p} {x \over L} {1 \over \sqrt{2 \pi L \sigma^2}}
\exp \left[ -{(x-\overline{x})^2 \over 2 L \sigma^2} \right]\;,
\label{PLx}
\ee
with
\be
\overline{x}=(p-q)L=-\delta_w L,~~~\sigma^2=4pq\;.
\label{xsigma}
\ee
The surviving probability is then given by:
\end{multicols}
\renewcommand{\theequation}{\Alph{section}\arabic{equation}}
\widetext
\noindent\rule{20.5pc}{.1mm}\rule{.1mm}{2mm}\hfill
\be 
P_{\rm surv}(\delta_w,L)=\int_0^{\infty} P_L(x)dx=\sqrt{2 \sigma^2
\over p^2 \pi L} \exp(-\overline{y}^2) \left\{ {1 \over 2}+
{\overline{y} \over 2} \sqrt{\pi} \exp(\overline{y}^2) \left[
1-\Phi(-\overline{y}) \right] \right\}\;,
\label{psurvgen}
\ee
\hfill\rule[-2mm]{.1mm}{2mm}\rule{20.5pc}{.1mm}
\begin{multicols}{2} 
\narrowtext
\noindent 
where
\be
\overline{y}={\overline{x} \over \sqrt{2L \sigma^2}}=-\delta_w \sqrt{L \over
2 \sigma^2}\;,
\label{ybar}
\ee
and $\phi{z}=2/\sqrt{\pi} \int_0^z \exp(-t^2) dt$ is the error
function\cite{gradstein}.

In the following we evaluate $P_{\rm surv}(\delta_w,L)$ in
eq(\ref{psurvgen}) in the different limits.

In the symmetric case $\delta_w=0$, $p=1/2$:
\be
P_{\rm surv}(\delta_w=0,L)= {1 \over \sqrt{8 \pi L}} \sim L^{-1/2}\;.
\label{psurv0}
\ee
For $\delta_w>0$, when $p<q$ and the walk has a drift towards the wall the
surviving probability has an exponential decay as $\overline{y} \to - \infty$:
\beqn
P_{\rm surv}(\delta_w>0,L)&=&\sqrt{2 \sigma^2 \over
p^2 \pi L} \exp(-\overline{y}^2)  {1 \over 4 \overline{y}^2} \nonumber\\
&\sim & L^{-1/2} \exp(-L/\xi_w) \xi_w/L\;,
\label{psurv+}
\eeqn

with a correlation length:

\be
\xi_w={2 \sigma^2 \over \delta_w^2}\;.
\label{wcorr}
\ee

Finally, for $\delta_w<0$, when $p<q$ and the walk is drifted from the wall
the surviving probability has a finite limit:

\be
P_{\rm surv}(\delta_w<0,L)=\sqrt{2 \sigma^2 \over
p^2 \pi L} \overline{y}=-{\delta_w \over p}\;.
\label{psurv-}
\ee

The size of {\it transverse fluctuations} of the adsorbing walk is given by:

\be
l_{\rm tr}(\delta_w,L)=\int_0^{\infty} P_L(x) x dx/P_{\rm surv}(\delta_w,L)\;,
\label{trfluct}
\ee
where
\end{multicols}
\widetext
\noindent\rule{20.5pc}{.1mm}\rule{.1mm}{2mm}\hfill
\be
\int_0^{\infty} P_L(x) x dx={2 \sigma^2 \over
p \sqrt{\pi}} \exp(-\overline{y}^2) \left\{ {\overline{y} \over 2}
+{\sqrt{\pi} \over 4}(2 \overline{y}^2+1) \exp(\overline{y}^2)
\left[ 1-\Phi(-\overline{y}) \right] \right\}\;.
\label{integ}
\ee
\hfill\rule[-2mm]{.1mm}{2mm}\rule{20.5pc}{.1mm}
\begin{multicols}{2} 
\narrowtext
\noindent 
In the symmetric limit $\delta_w=0$:
\be
l_{\rm tr}(\delta_w=0,L)=\sqrt{8 \pi L} \sim L^{1/2}\;.
\label{trfluct0}
\ee
For $\delta_w>0$ the transverse fluctuations in leading order are independent
of $L$:
\be
l_{\rm tr}(\delta_w>0,L)={2 \sigma^2 \over \delta_w}\;,
\label{trfluct+}
\ee
while for $\delta_w<0$, when there is a drift of the walk from the wall the
transverse fluctuations grow linearly with $L$:
\be
l_{\rm tr}(\delta_w<0,L)=\delta_w L\;.
\label{trfluct-}
\ee
The {\it maximal value of the transverse fluctuations}
$l_{\rm tr}^{max}(\delta_w,L)$ for $\delta \le 0$ are in the same order of
magnitude as their average values in eqs(\ref{trfluct0}) and
(\ref{trfluct-}). However for $\delta>0$ the maximal value is
generally larger, then the average one in eq(\ref{trfluct+}). In this
case $l_{\rm tr}^{max}(\delta_w>0,L)$ is determined by a {\it rare event},
in which a large fluctuation of positive steps is followed by a drift
process towards the average behavior. If the number of steps in the
drift process is $\alpha L$, where $0<\alpha<1$, then
\be
l_{\rm tr}^{max}(\delta_w>0,L)=\alpha L \delta_w\;.
\label{trfluctmax}
\ee

\end{multicols}
\end{document}